\documentclass[traditabstract]{aa}

\usepackage{graphicx}
\usepackage{txfonts}
\usepackage{natbib}
\usepackage[colorlinks=true,citecolor=blue]{hyperref}
\usepackage{xspace}

\usepackage[T1]{fontenc}

\bibpunct{(}{)}{;}{a}{}{,}

\newcommand{\MJup}{M$_{\mathrm{Jup}}$\xspace}

\newcommand{\teff}{T$_{e\!f\!f}$\xspace}
\newcommand{\logg}{log~\emph{g}\xspace}
\newcommand{\mic}{$\mu$m\xspace}
\newcommand{\as}{\hbox{$^{\prime\prime}$}\xspace}
\newcommand{\lsd}{\hbox{$\lambda/D$}\xspace}

\begin{document}

\title{Apodization in high-contrast long-slit spectroscopy}
\subtitle{Closer, deeper, fainter, cooler}

\author{A. Vigan\inst{1,2} \and M. N'Diaye\inst{2,3} \and K. Dohlen\inst{2}}

\institute{Astrophysics group, School of Physics, University of Exeter, Stocker Road, Exeter EX4 4QL, United Kingdom \\
\email{\href{mailto:arthur@astro.ex.ac.uk}{arthur@astro.ex.ac.uk}}
\and
Aix Marseille Universit\'e, CNRS, LAM (Laboratoire d'Astrophysique de Marseille) UMR 7326, 13388, Marseille, France
\and
Space Telescope Science Institute, 3700 San Martin Drive, Baltimore MD 21218, USA
}

\date{Preprint version: 22 May 2013}

\abstract{
The spectroscopy of faint planetary-mass companions to nearby stars is one of the main challenges that new-generation high-contrast spectro-imagers are going to face. However, the high contrast ratio between main-sequence stars and young planets makes it difficult to extract a companion spectrum that is not biased by the signal from the star. In a previous work we demonstrated that coupling long-slit spectroscopy (LSS) and classical Lyot coronagraphy (CLC) to form a long-slit coronagraph (LSC) allows low-mass companions to be properly characterized when combined with an innovative a posteriori data analysis methods based on the spectral deconvolution (SD). However, the presence of a slit in the coronagraphic focal plane induces a complex distribution of energy in the Lyot pupil plane that cannot be easily masked with a binary Lyot stop, creating strong diffraction residuals at close angular separation. To alleviate this concern, we propose to use a pupil apodization to suppress diffraction, creating an apodized long-slit coronagraph (ALSC). We show that this concept allows to look closer from the star, at deeper contrast, which enables the characterization of fainter substellar companions. After describing how the apodization was optimized, we demonstrate its advantages with respect to the CLC in the context of SPHERE/IRDIS LSS mode at low resolution with a 0.12\as slit and 0.18\as coronagraphic mask. We performed different sets of simulations with and without aberrations, and with and without a slit to demonstrate that the apodization is a more appropriate concept for LSS, at the expense of a significantly reduced throughput (37\%) compared to the LSC. Then we performed detailed end-to-end simulations of the LSC and the ALSC that include realistic levels of aberrations to obtain several datasets representing 1~h of integration time on stars of spectral type A0 to M0 located at 10~pc. We inserted the spectra of planetary companions at different effective temperatures (\teff) and surface gravities (\logg) into the data at angular separations of 0.3\as to 1.5\as and with contrast ratios from 6 to 18~mag. Using the SD method to subtract the speckles, we show that the ALSC brings a gain in sensitivity of up to $\sim$3~mag at 0.3\as over the LSC and that both concepts are essentially equivalent for separations larger than 0.5\as. The gain at small separation is the result of suppressing of the bright Airy rings that are difficult to estimate at very small angular separations because of the PSF chromaticity. The improved sensitivity is confirmed by extracting the simulated companions spectra from the data and comparing them to libraries of models to determine their \teff and \logg. Using a restoration factor that quantitatively compares the input and output spectra, we show that the ALSC data systematically leads to better quality spectra below 0.5\as. In terms of \teff, we demonstrate that at small angular separations the limit with the ALSC is always lower by at least 100~K, inducing an increase in sensitivity of a factor up to 1.8 in objects' masses at young ages. Finally, for the determination of \logg, we show that the ALSC provides a less biased estimation than the LSC.
}


\maketitle

\section{Introduction}
\label{sec:introduction}

The number of imaged planetary-mass companions to main-sequence stars is increasing steadily thanks to significant improvements in high-contrast instrumentation and data analysis methods, as well as the execution of large-scale surveys targeting young nearby stars to maximize detection chances \citep[e.g.][]{chauvin2010,vigan2012b}. When it comes to characterizing the atmospheric (composition, cloud properties, etc.) and physical (effective temperature -- \teff, surface gravity -- \logg, radius, mass, luminosity) properties of these low-mass companions, it becomes essential to obtain high signal-to-noise ratio (S/N) emission spectra that can be compared to libraries of synthetic or empirical atmosphere models and that eventually lead to a better calibration of the models.

To reach the very high contrasts needed to detect young giant planets, forthcoming instruments (e.g. \citealt{macintosh2008,beuzit2008}) will couple extreme adaptive optics systems \citep{angel1994,stahl1995} and high-efficiency coronagraphs (see \citealt{guyon2006} for a review), the combination of which is essential to achieve very high correction of the atmospheric turbulence in the near-infrared and to obtain an optimal extinction of the star and the diffraction. Back-end scientific instruments behind these systems will provide an advanced means of characterizing the detected planets, such as diffraction-limited integral field spectroscopy (IFS, e.g. \citealt{antichi2009}), long-slit spectroscopy (LSS, e.g. \citealt{vigan2008}), or polarimetry (e.g. \citealt{joos2011}).

In our previous work \citep{vigan2008} we studied the combination of LSS with a classical Lyot coronagraph (CLC), thereby creating a long-slit coronagraph (LSC), for the instrument SPHERE/IRDIS (InfraRed Dual Imaging Spectrograph, \citealt{dohlen2008}). In high-contrast applications, LSS has some significant advantages over IFS, since the use of lenslet arrays in the latter generally limits the field of view ($\lesssim$~2\as), the spectral bandwidth ($\lesssim$~1~\mic) and/or the resolution ($\lesssim$~50). Such limitations do not affect LSS, but there are others that are intrinsic to the use of a slit combined with a Lyot coronagraph in the same focal plane. In imaging with a Lyot coronagraph, the circular pupil with central obscuration and spiders produces a highly axisymmetrical energy distribution in the Lyot pupil plane that is easily masked with a circular Lyot stop (with oversized central obscuration and spider masks). In the case of LSC, the introduction of a slit in the coronagraphic plane completely redistributes the energy in the Lyot pupil plane such that the use of an axisymmetric Lyot stop becomes useless, as we demonstrate later on.

There are two possible solutions to this problem. The first one is to introduce a very complex, non-axisymmetric Lyot stop that would significantly decrease the instrument throughput. The second one is to get rid of the need for a Lyot stop with the use of an optimized apodization that will concentrate the stellar light within the central peak, rather than eliminating it. Such an apodization is obtained either by multiplication of the entrance pupil with a smooth/continuous function \citep{nisenson2001,gonsalves2003,aime2005b} or by using a shaped/binary pupil function \citep{kasdin2003,vanderbei2003b,vanderbei2003,vanderbei2004,aime2005b}. In this case, the diffraction rings of the stellar PSF are considerably reduced, resulting in higher contrast in the search area, but an opaque coronagraphic mask is still required to mask out the brighter central peak. A concept where it is not possible to introduce a Lyot stop in the optical path -- for optomechanical or physical reasons -- has already been proposed for the instrument GTC/FRIDA, and was dubbed the Stop-Less Lyot Coronagraph \citep[SLLC, ][]{ndiaye2007,ndiaye2008}.

In \citet{vigan2008} we showed that the use of the LSC resulted in strong diffraction residuals at very small angular separations (0.2\as--0.5\as), which decreases the characterization capabilities in this very interesting range where the next-generation high-contrast imagers are supposed to give their full potential. That is why the use of a device that truly suppresses the diffraction residuals over a wide spectral range appears crucial for obtaining LSS data limited by the speckle noise at small angular separations. In the present work, we thus compare two approaches: the LSC and the combination of the SLLC with LSS, which produces an apodized long-slit coronagraph (ALSC).

First, in Sect.~\ref{sec:stop_less_lyot_coronagraph} we describe the SLLC and how it compares with an optimized CLC in imaging and when a slit is introduced in the coronagraphic focal plane. Then in Sect.~\ref{sec:high_contrast_spectroscopy_simulations} we present the end-to-end simulations that we have performed to produce realistic high-contrast data for the LSC and the ALSC in an instrument similar to SPHERE/IRDIS. In Sect.~\ref{sec:speckle_noise_attenuation_sd} we compare the two concepts in terms of speckle noise attenuation with the data analysis method described in \citet{vigan2008}, and finally in Sect.~\ref{sec:spectral_extraction} we demonstrate the gain of using ALSC in terms of spectral extraction and characterization of planetary-mass companions.

\section{Stop-less Lyot coronagraph}
\label{sec:stop_less_lyot_coronagraph}
 
\subsection{Interest of the SLLC}
\label{sec:interest_sllc}

Pupil apodization suppresses the diffraction pattern of the telescope by concentrating most of the energy of the PSF inside a central peak rather than inside the Airy rings. The use of a focal plane opaque mask is, however, still required as an antisaturation and stray-light control device. As long as this mask fully covers the central peak of the apodized PSF, diffraction at the mask edge will be minimal, keeping the Lyot stop from being imperative. In the following, we use the term stop-less Lyot coronagraph (SLLC), previously introduced by \citet{ndiaye2007,ndiaye2008}, to refer to the combination of such a pupil apodization and a focal plane mask.

In the IFS mode of FRIDA, the forthcoming spectro-imager for the adaptive optics system of the \emph{Gran Telescopio Canarias} \citep[GTC,][]{lopez2007}, upgrade paths for high-contrast imaging using Lyot or phase spots and pupil masks have been considered in the design. APLC was one of the coronagraphic configurations proposed for FRIDA \citep{ndiaye2007}. Unfortunately, optomechanical constraints make the implementation of a fully fledged APLC with an apodizer, focal plane mask, and Lyot stop impossible. A pupil plane suitable for introducing an apodizer is available, but the following focal plane will contain the image slicer optics associated with the integral field spectrograph. Even though an opaque focal plane mask could conceivably be introduced in the plane of the slicers, the following pupil image would be spread out due to diffraction at the slicers, and the use of a Lyot stop in the following pupil plane would therefore be inefficient. Also, mechanical constraints of the FRIDA concept makes high-precision positioning of a Lyot stop impossible.
 
We note that similar constraints exist in the LSS mode of SPHERE/IRDIS. Here, the slit, which includes a stellar blocker in the center, is located in the coronagraphic focal plane \citep{dohlen2008}. The diffractive spread of the pupil in the direction perpendicular to the slit has been observed well in simulations, and although a Lyot stop is present in the spectrograph pupil, its efficiency is limited. The use of a suitable apodizer, as proposed by the SLLC concept, may provide improved efficiency in the mode considered for these instruments since its transmission function is optimized to minimize starlight leaks inside the Lyot plane. This apodizer could also represent an interesting solution for E-ELT/EPICS, since apodization is currently considered as a baseline for diffraction suppression \citep{verinaud2010}.

The SLLC concept includes important advantages over most classical coronagraphs. First, provided the apodizer transmission is spectrally gray, the apodizer efficiency is perfectly achromatic. Second, the optical system can be significantly simplified by omitting of intermediate image and pupil planes. Third, optical surfaces located in the focal plane can be avoided. In conjunction with a Lyot stop, they have been shown to severely limit the efficiency of spectrally based speckle calibration techniques \citep{verinaud2010}.
  
Several pupil plane amplitude masks have already been proposed, as mentioned in the review of apodization designs. We focus our work here on continuous rather than binary apodizers. Simplicity of fabrication and better throughput are often referred to as the main advantages of binary over smoothed apodizers. However, in its Radon approach to compare binary and smoothed apertures, \citet{aime2005b} underlines that binary pupil apodizers spread light uselessly in some directions contrary to the continuous apodizations. Most of these smoothed apodizers were designed for unobstructed apertures. Nevertheless, 8--10~m class telescopes usually have an on-axis secondary mirror, the shadow of which is projected on the primary mirror, leading to a pupil with central obscuration. In the following, we describe the computation of the SLLC apodizer, considering the possible presence of a central obscuration.

\subsection{Generation of the apodization}

\begin{figure}
  \centering
  \includegraphics[width=0.5\textwidth]{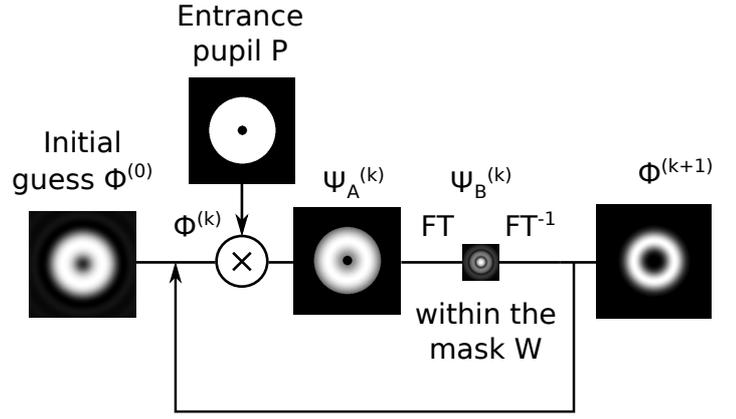}
  \caption{Schematic illustration of the basic algorithm for computing the SLLC apodization.}
  \label{fig:algo}
\end{figure}

For calculating apodizer transmission profiles, \citet{gonsalves2003} developed an algorithm that optimizes the design of a classical pupil apodizer to reduce the bright diffraction rings of a star image. The general method can theoretically be used for any telescope aperture geometry, but in their paper the authors did not address the case of centrally obstructed aperture. We computed the SLLC apodization, adapting this algorithm to the case of the VLT pupil. In the following, we describe the scheme of the algorithm, summarized graphically in Figure \ref{fig:algo}. For the sake of clarity, we omit the spatial coordinate $\textbf{r}$, its modulus $r$ and the wavelength $\lambda$ in the equation below, and $\mathcal{F}$ denotes the Fourier transform operator. 

$\Phi^{(k)}$ denotes the apodizer transmission function at iteration $k$. The complex amplitude of the field in the entrance pupil plane $\Psi^{(k)}_A$ can be expressed as follows:
\begin{equation}
\Psi^{(k)}_A = P\,\Phi^{(k)}\,,
\end{equation}
in which $P$ defines the telescope aperture shape.

The PSF is the squared Fourier transform of the previous expression. Since it deals with apodization, we aim to concentrate most of the starlight in the core and first two or three bright rings of the PSF, thus reducing the luminosity in its following bright rings so as to ease the observation of planetary companions in the search area. To reach this goal, we filter the previous amplitude with a mask $W$ whose transmission is non-null inside the mask area and $0$ outside. Therefore, the complex amplitude in the following focal plane $\Psi^{(k)}_B$ can be written as
\begin{equation}
\Psi^{(k)}_B = \mathcal{F} \left [\Psi^{(k)}_A \right ] \times W\,.
\end{equation}
Since the Fourier transform calculus of $\Psi^{(k)}_A$ is only required inside the mask area, we can perform the semi-analytical method using the matrix Fourier transform proposed by \citet{soummer2007b} instead of the traditional fast Fourier transform to achieve a faster computation of this algorithm.

Finally, the apodizer transmission function at iteration $k+1$ is given by 
\begin{equation}
\Phi^{(k+1)} = \mathcal{F} \left [\Psi^{(k)}_B \right ]\,.
\end{equation}
A filter $\Omega$ can be introduced here to smooth the apodizer shape as mentioned by \citet{gonsalves2003}. In our final computation of the apodizer, we kept this function equal to $1$ because various tests showed that no significant improvement on the performance could be obtained by varying this function.

These operations are repeated iteratively until reaching a satisfying apodizer. An initial apodizer shape $\Phi^{(0)}$ can also be introduced in the algorithm as suggested by \citet{gonsalves2003} but the starting point was not important in our simulations to obtain our solution. 

\begin{figure}
  \centering
  \includegraphics[width=0.5\textwidth]{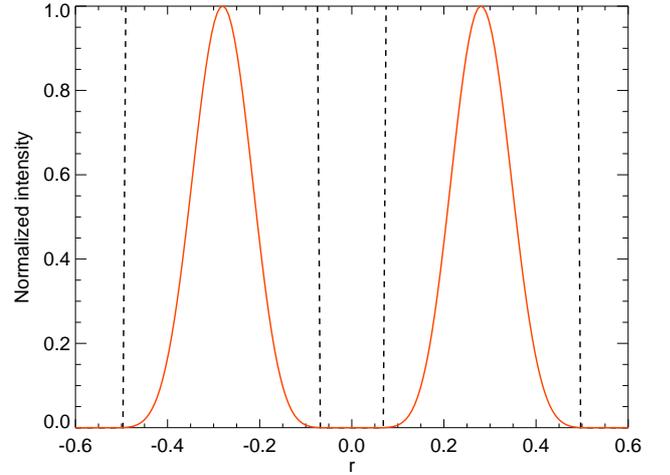}
  \caption{Radial profile of the SLLC intensity apodization optimized for SPHERE/IRDIS and an opaque mask of radius 4.53\,$\lambda_0/D$. The throughput is 37\%. The dashed lines delimit the pupil of the VLT.}
  \label{fig:sllc_apod}
\end{figure}

As an illustration, we compute the SLLC apodizer for SPHERE/IRDIS, considering an opaque mask of 0.18\as angular radius on sky, or equivalently 4.53\,$\lambda_0/D$ at $\lambda_0=$1.59\,$\mu$m, see Figure \ref{fig:sllc_apod}. VLT spiders were not considered in the optimization to avoid an asymmetric apodization shape. 
 
\subsection{Comparison SLLC/CLC}

\begin{figure*}
  \centering
  \includegraphics[width=0.49\textwidth]{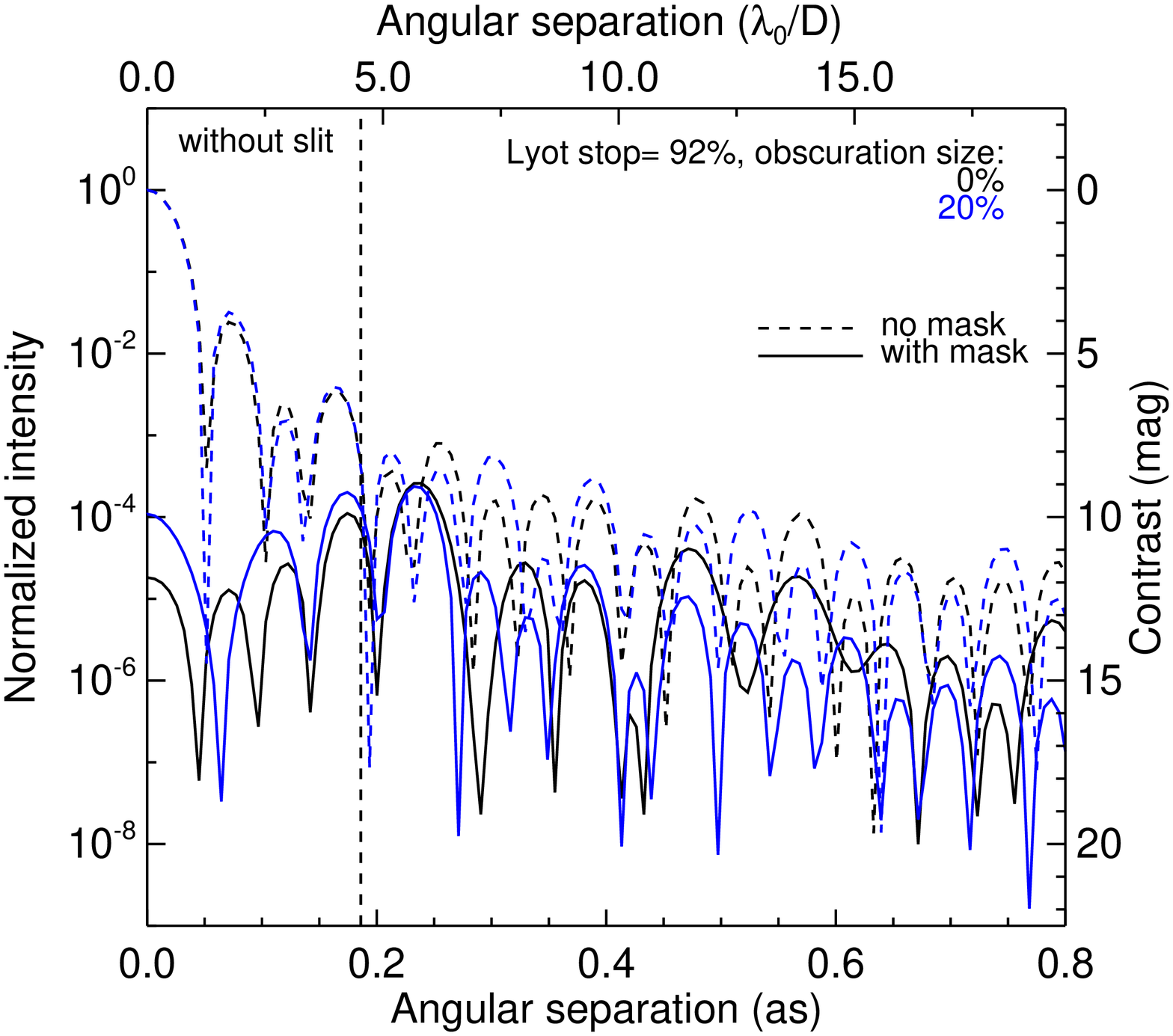}
  \includegraphics[width=0.49\textwidth]{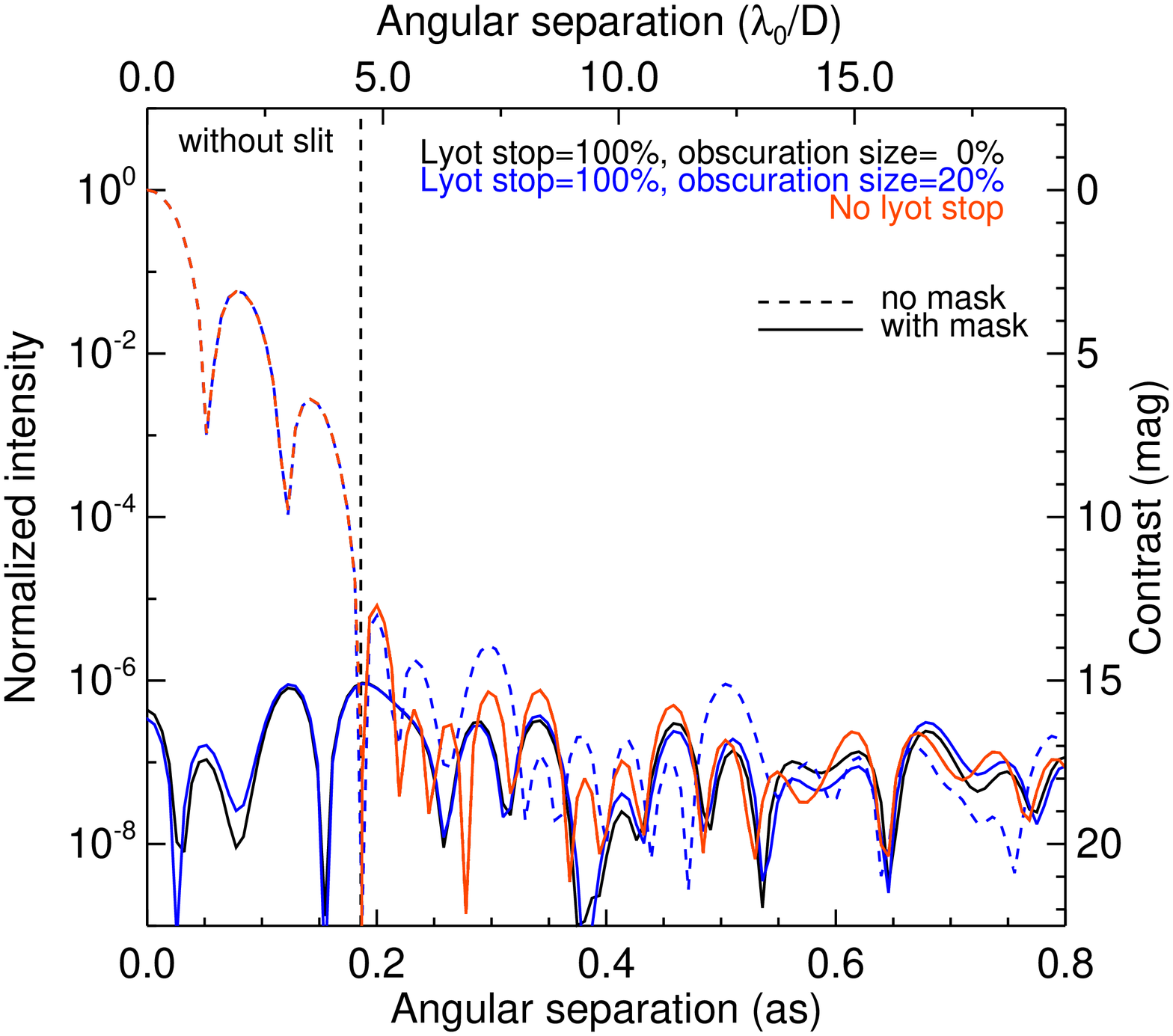}
  \caption{Radial intensity profiles of the coronagraphic images with CLC (left) or SLLC (right) for different Lyot stop configurations (see legend in figures for description) in imaging mode (no slit). The vertical line represents the opaque mask radius.}
  \label{fig:contrast_noslit}
\end{figure*}

\begin{figure*}
  \centering
  \includegraphics[width=1.0\textwidth]{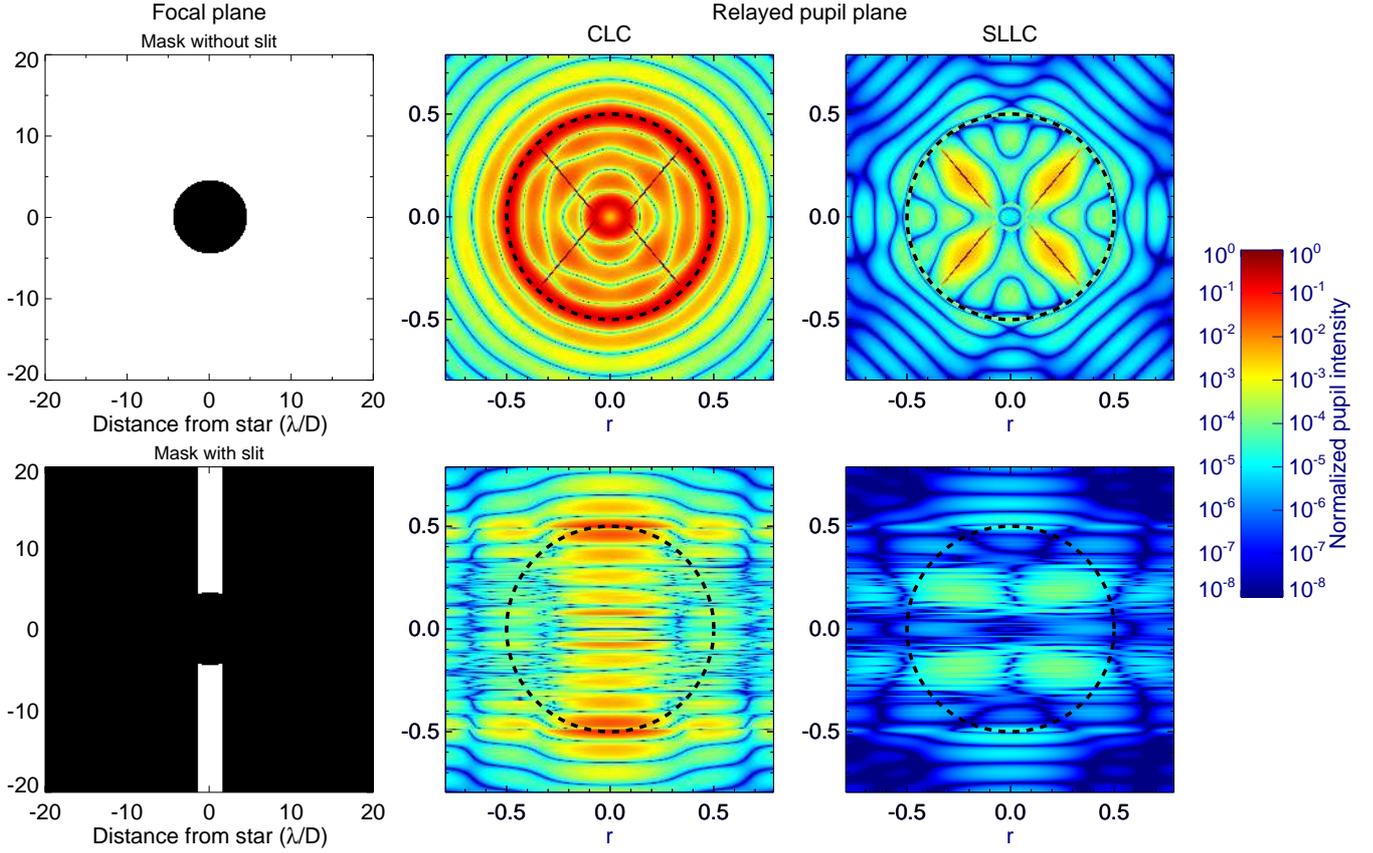}
  \caption{Energy distribution in the relayed pupil-planes for both concepts in imaging and slit modes. From left: focal plane mask transmission function, light distribution in the relayed pupil plane with CLC (middle), and SLLC (right). Top and bottom frames correspond to imaging and slit modes, respectively. The size of the entrance pupil in the relayed pupil planes is plotted with a thick dashed circle.}
  \label{fig:pupil_images}
\end{figure*}

\begin{figure*}
  \centering
  \includegraphics[width=0.49\textwidth]{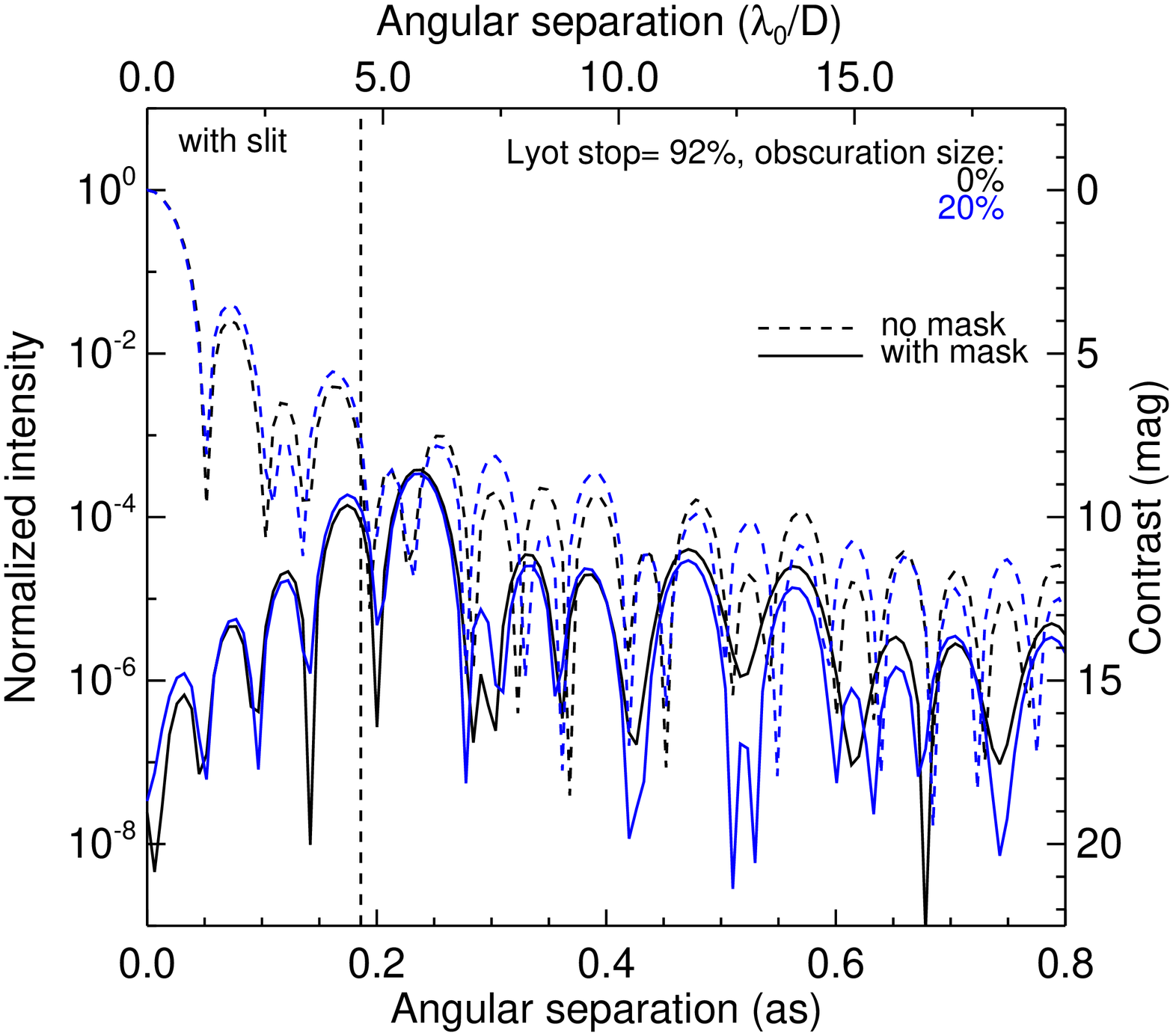}
  \includegraphics[width=0.49\textwidth]{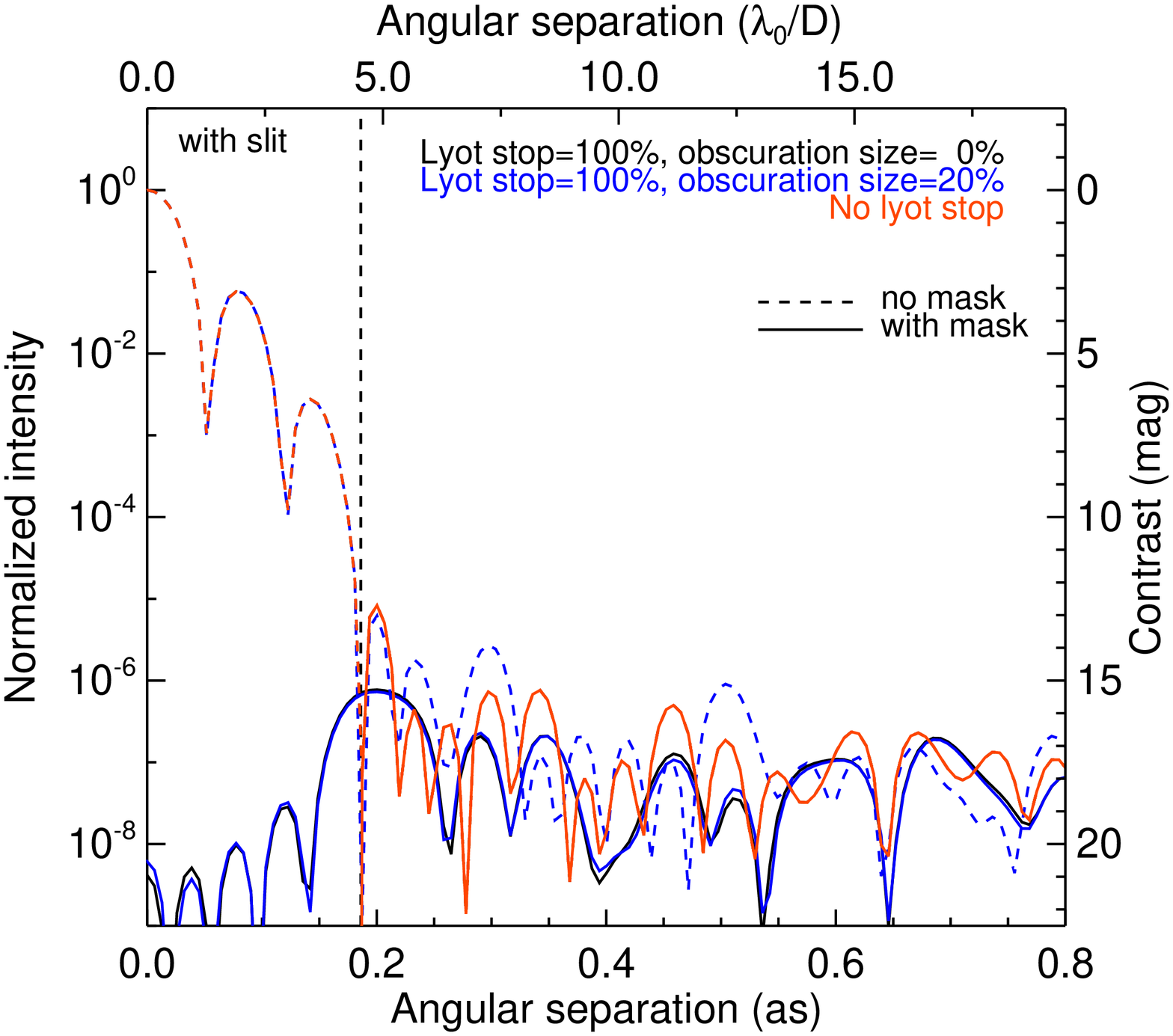}
  \caption{Same as Fig. \ref{fig:contrast_noslit} but here, in the presence of a slit. The profiles are obtained with an average of the coronagraphic images along the slit width.}
  \label{fig:contrast_slit}
\end{figure*}

\begin{figure*}
  \centering
  \includegraphics[width=0.49\textwidth]{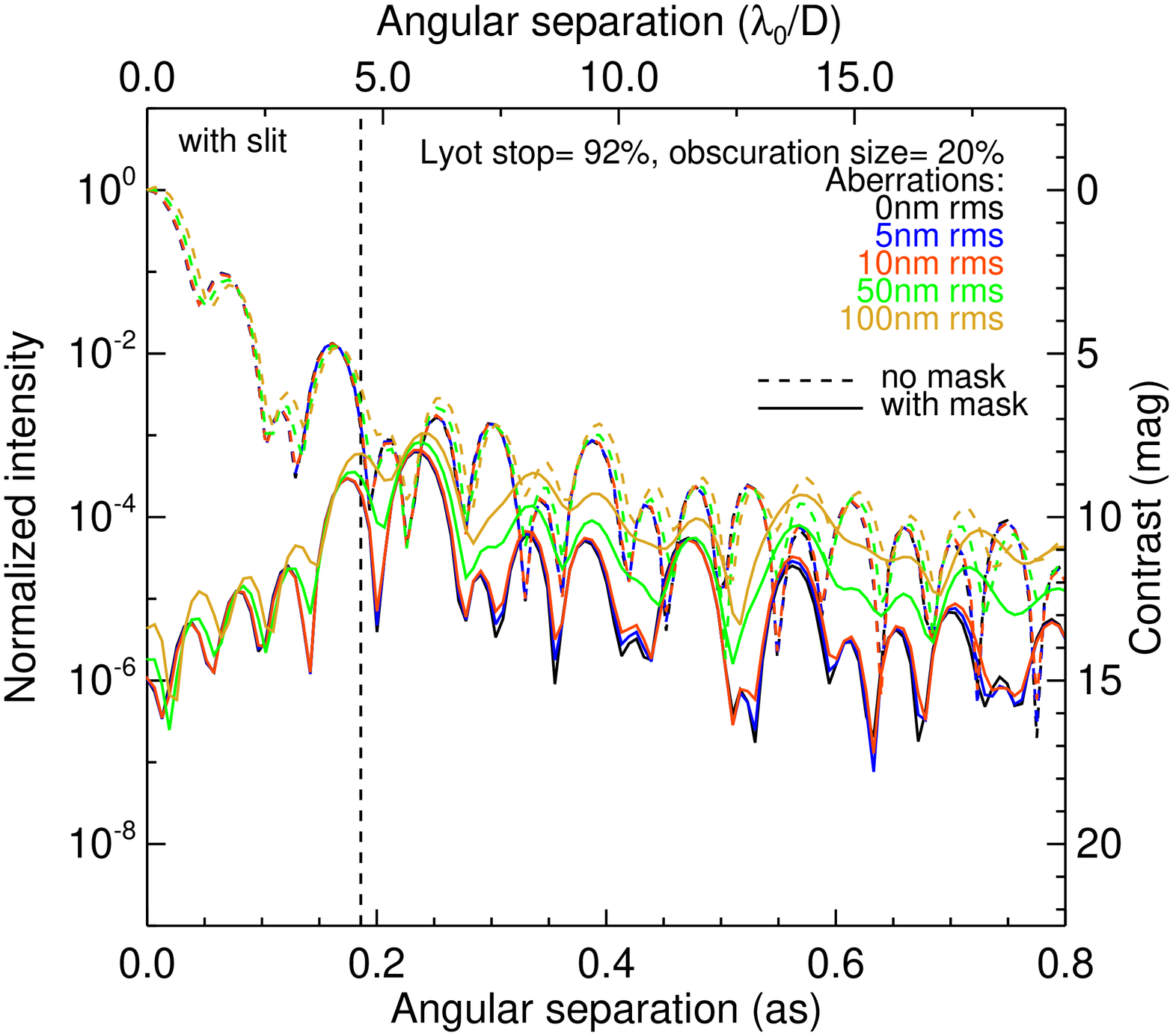}
  \includegraphics[width=0.49\textwidth]{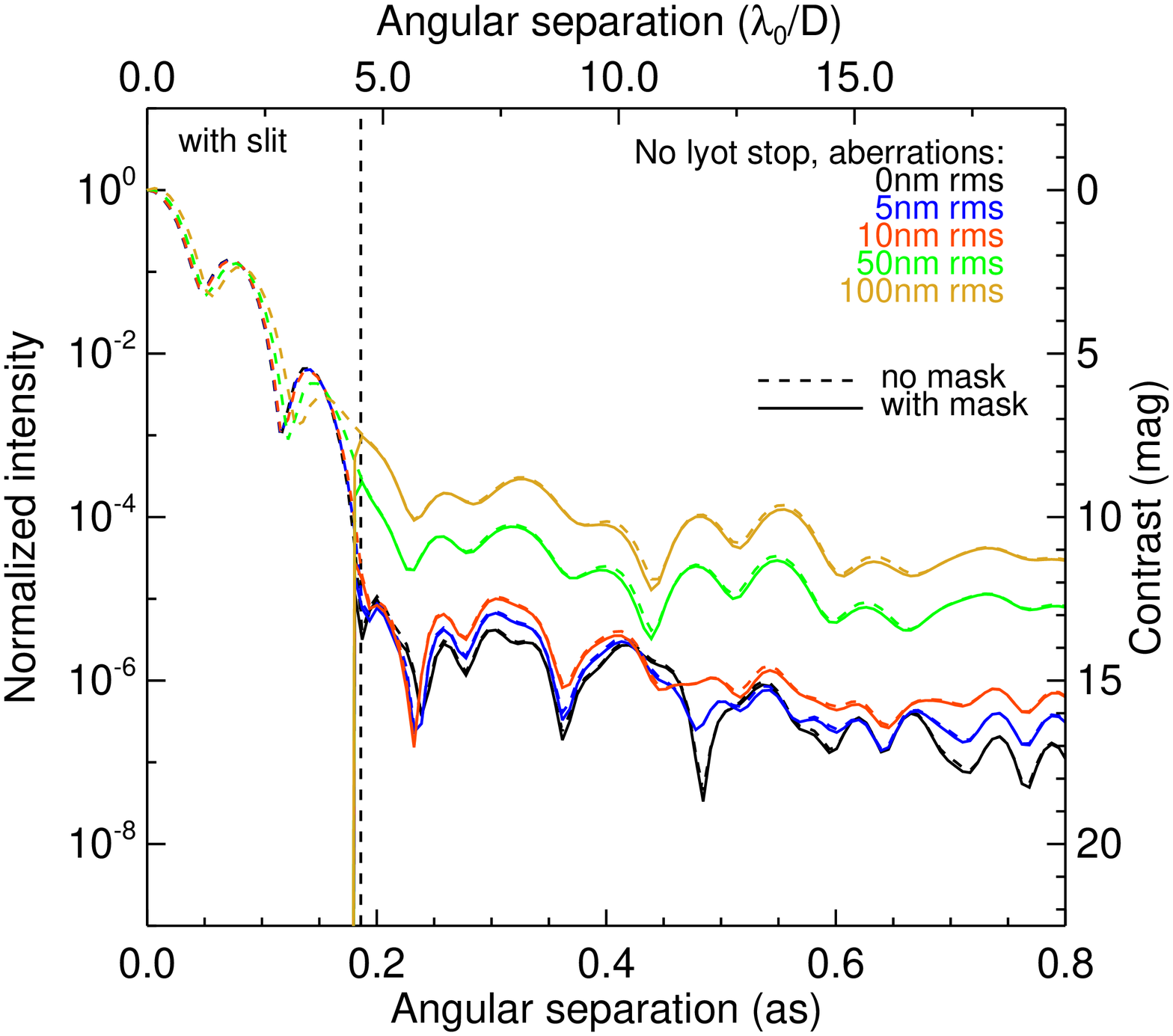}
  \caption{Average intensity profile of the coronagraphic images along the slit width with CLC (left) or SLLC for different values of static aberrations (see values in figures) with $\nu^{-2}$-power spectral density. The vertical line represents the opaque mask radius.}
  \label{fig:contrast_aberrations}
\end{figure*}

We now estimate the performance of SLLC and compare it to that of CLC, first addressing the imaging mode and then the slit mode. In the following numerical simulations, we consider the VLT pupil with its spider vanes and a central obscuration of 14\% of the telescope diameter, a 0.18\as-radius focal plane opaque mask as in SPHERE/IRDIS for both concepts, and a 0.12\as-width slit, and we assume observations in H-band ($\lambda_0=1.594\,\mu$m). Lyot stop refers to the outer pupil stop, while the percentage for inner and outer stops indicates the fraction of the entrance pupil size.

In the presence of an inner stop, a spider stop is introduced with a size corresponding to that of the re-imaged spider vanes. For a given concept and pupil stop configuration, the intensity profiles of the coronagraphic images are normalized with respect to the intensity peak of the images without an opaque mask. The intensity level is estimated by averaging the coronagraphic profile over an annular ring of 0.04\as ($\lambda_0/D$) width at a given angular separation. As a first step, no aberrations are introduced in the system.

We plot the radial intensity profiles of the coronagraphic images with CLC and SLLC in Fig. \ref{fig:contrast_noslit} for the imaging mode and we consider different relayed pupil stop configurations for each concept. To allow further comparisons, the curves are extracted from a direction outside the orientation of the spider diffraction effects, following that of the slit in slit mode.

In the right-hand plot of Fig. \ref{fig:contrast_noslit}, SLLC without pupil stop reaches an intensity level of $2.2 \times 10^{-7}$ at 0.25\as and slightly lower values with a pupil stop having null (black) or 20\% (blue) central obscuration. In the latter configurations, the residual light within the re-imaged opaque mask is due to the pupil stop diffraction. Although slight performance improvements are observed with a relayed pupil stop, our concept retains a good ability to remove starlight without this device.

The left-hand plot of Fig. \ref{fig:contrast_noslit} displays the performance of CLC for different central obscuration stop sizes and an outer pupil stop with a size of 92\% of the entrance pupil diameter as in the SPHERE/IRDIS LSS mode to mostly eliminate the diffraction due to the focal plane mask. The contrasts in the presence of a centrally obstructed pupil stop prove larger than in the case of a clear pupil stop, underlining the interest in the inner stop for removing the diffraction due to the entrance pupil obscuration. Additional tests showed little impact of the inner stop size on the overall coronagraphic performance for values beyond 20\%, which remained larger than the 14\% entrance obscuration diameter. A common intensity level of $7.7 \times 10^{-5}$ is reached at 0.25\as, which proves to be higher than the SLLC value and thus shows a favorable comparison for our concept with respect to CLC with any pupil stop configuration.

In slit mode, the addition of a slit to the coronagraphic mask in the focal plane leads to light diffraction in the direction perpendicular to the slit in the relayed pupil plane for both SLLC and CLC concepts, see Fig. \ref{fig:pupil_images} bottom frames. Compared to the imaging mode (top frames), the light distribution is no longer axisymmetric, making a circular pupil stop useless.  In addition, there is much less residual light left in both imaging and slit modes with SLLC than with CLC, underlining the efficiency of our concept.

The uselessness of a pupil stop in slit mode is confirmed with the analysis of the coronagraphic intensity profiles that are generated by averaging the images along the slit width (see Fig. \ref{fig:contrast_slit}). Equivalent performance is obtained between the different pupil stop configurations for each concept, and no major difference is observed compared to the imaging case: intensity levels of $2.2 \times 10^{-7}$ and $2.0 \times 10^{-4}$ are achieved respectively with SLLC and CLC at 0.25\as. Our concept performs better than CLC while offering promising contrast gains in slit mode with no aberrations.

Is this assertion still valid in the presence of quasi-static aberrations? To answer this question, we numerically simulated both SLLC and CLC in the presence of wavefront errors, assuming one phase screen in the entrance pupil plane with a $\nu^{-2}$-power spectral density where $\nu$ denotes the spatial frequency. Figure~\ref{fig:contrast_aberrations} displays coronagraphic intensity profiles for different amounts of aberration. For small aberrations (<10~nm RMS), SLLC is more sensitive to phase errors than CLC owing to its better ultimate performance, but it still remains more efficient than CLC. At higher aberrations ($\geq$ 50~nm RMS), SLLC and CLC contrasts prove almost equivalent, as expected since diffraction suppression systems are all equivalently inefficient for large aberrations.

\section{High-contrast spectroscopy simulations}
\label{sec:high_contrast_spectroscopy_simulations}

For applications to LSS, we compare the LSC and the ALSC using more detailed end-to-end simulations of realistic high-contrast data for both designs. The overall instrumental design is that of VLT/SPHERE with a coronagraphic long-slit mask in the IRDIS science module. As in our previous work, we use a diffractive code written in IDL based on the CAOS (code for adaptive optics systems) problem-solving environment \citep{carbillet2004} with a package specifically developed for VLT/SPHERE simulations \citep{carbillet2008}.

The code differentiates the simulation of the instrument (Sect.~\ref{sec:simulation_instrument}), which generates the normalized data, and the simulation of the photometry (Sect.~\ref{sec:simulation_photometry}), which use the normalized data to generate realistic observations in terms of photometry and noise.

\subsection{Simulation of the instrument}
\label{sec:simulation_instrument}

The code handles the simulation of the full instrument taking multiple sources of optical aberrations into account. The content and details of the SPHERE simulation package have been described previously in \citet{carbillet2008,carbillet2011}, but we recall below the important steps and details of the simulations.

The first part of the simulation is the high-order AO system (40$\times$40 actuators), which produces the filtered instantaneous atmospheric residuals. The working parameters of the SPHERE AO system are all described in \citet{fusco2006} and \citet{petit2008}, and the ones of interest for the current simulations are

\begin{itemize}
\item the $V$-band star magnitude, which will partly define the overall AO system performance. We simulate data for $V = 4$ (bright star, optimal AO performance) and $V = 8$ (dim star, average AO performance);

\item the seeing of the observations, which is another driving parameter for the AO performance. A value of 0.8\as is used in our simulations, which represents average observing conditions;

\item the instrumental jitter is set to 3~mas, which is the baseline for the SPHERE instrument. This jitter is chosen randomly at each realization of an AO-filtered phase screen and is introduced as tip/tilt term into the phase screens.
\end{itemize}

\noindent The simulation assumes that the wavefront sensing and observations are performed in $R$-band ($\lambda = 650$~nm) and in the near-IR ($Y$--$K$), respectively. The residual errors of the SPHERE AO system for a $V = 8$ star and 0.85\as seeing are on the order of 90~nm~RMS.

The second part of the code handles the static aberrations of the common path optics of the instrument, from the telescope mirror to the coronagraph plane. The aberrations are introduced as phase screens explicitely included in the simulation package or generated from a precalculated power spectrum density (PSD). The common path optics include the following terms \citep{boccaletti2008}: the measured phase maps of the VLT-UT3 M1, M2, and M3 mirrors, for which low frequencies have been filtered out to take the effect of the AO system into account (28.6~nm RMS), the aberrations introduced by the various optics (34.5~nm~RMS), the calibration of the AO system (7.4~nm~RMS), the propagation (4.7~nm~RMS), and the beam shift (8~nm~RMS). In these simulations, the full Fresnel propagation of the wavefront is not considered.

The third part of the code handles the simulation of the coronagraph and IRDIS in its LSS mode. For the coronagraph, there are three important parameters that allow simulation of different coronagraph designs: the pupil apodization, the coronagraphic mask, and the Lyot stop. For both the LSC and ALSC, we use a 186-mas radius opaque coronagraph mask (4.53$\lambda/D$ at 1.6~\mic) at the center of a 120~mas wide slit. The apodization is clear for the LSC and is set to the apodization function calculated in Sect.~\ref{sec:stop_less_lyot_coronagraph} for the ALSC. The Lyot stop is a circular aperture undersized (92\%) with respect to the entrance pupil for the LSC, and there is no Lyot stop for the ALSC. None of the Lyot stops include spider-vane masking because LSS observations in fixed field will result in rotation of the entrance pupil for an alt-az telescope. The entrance pupil is set to a circular aperture with central obstruction and spider orientation matching those of the VLT. The code also introduces the aberrations downstream of the coronagraph ($\sim$30~nm~RMS in LSS), the chromatic pointing offset introduced by the atmospheric dispersion compensator (ADC), and a temporal pointing offset term (0.35~mas~RMS). There is no temporal evolution of the offset other than the 3~mas jitter taken into account in the simulation of the AO system. Since we are simulating spectroscopic data, the spectral resolution is defined by the number of simulated wavelengths. We simulate a low-resolution spectroscopy mode covering $Y$- to $K_{\mathrm{s}}$-band (0.95--2.3~\mic), sampled by 100 wavelengths, at an average resolution $R \simeq 60$.

Taking all the instrumental effects into account, the level of quasi-static aberrations reaches $\sim$55~nm~RMS. Additionally, if we include the level of AO residuals, we reach a level of $\sim$106~nm~RMS. To have independent data sets, i.e. data with different realizations of the quasi-static aberrations, we perform five series of simulations with different random seeds for the quasi-static aberrations. Each series includes simulated data for both coronagraph designs and for two AO guide stars with $V = 4$ and $V = 8$. For each simulation, 100 wavelengths are simulated to obtain images of the slit covering the full $Y$ to $K_{\mathrm{s}}$ spectral range. Finally, given the large amount of time needed to perform each of these simulations, the temporal evolution of the system cannot be fully simulated (continuous variation of the aberrations, rotation of the entrance pupil, etc.). The only temporal effects effectively considered are the evolution of the AO-filtered atmospheric residuals and the instrumental jitter. For each simulation, the final data are obtained by averaging images produced from 100 decorrelated atmospheric phase screens, which produces a smooth stellar halo over which the speckles induced by the static instrumental aberrations are superimposed.

\subsection{Simulation of the photometry}
\label{sec:simulation_photometry}

\begin{figure}
  \centering
  \includegraphics[width=0.5\textwidth]{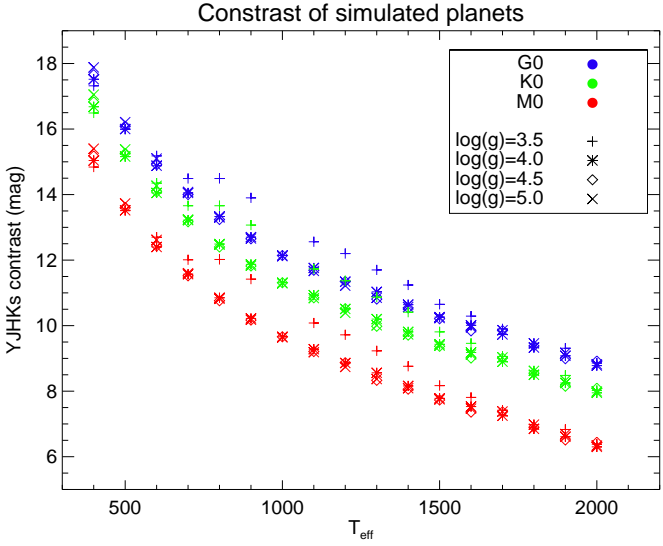}
  \caption{Average $YJHK_{\mathrm{s}}$ contrast of all simulated planets around G0 (blue), K0 (green), and M0 (red) stars at 10~pc as a function of \teff. Three differents colors are used for the three stellar spectral types, while different symbols are used for the \logg value.}
  \label{fig:contrast_all_planets}
\end{figure}

\begin{figure}
  \centering
  \includegraphics[width=0.45\textwidth]{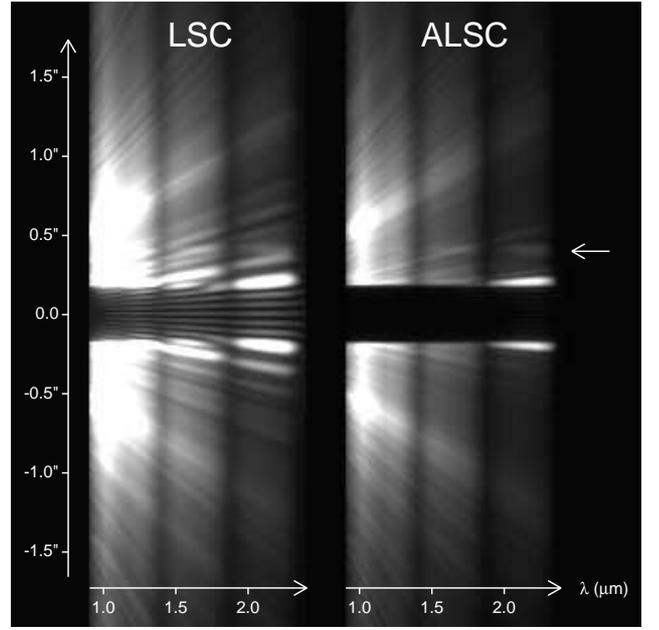}
  \caption{Simulated data with the LSC (left) and ALSC (right) designs representing a 1-hour total integration time on a G0 star at 10~pc. A fake companion (\teff = 1400~K, \logg = 5.0) has been introduced at a separation of +0.4\as from the star (left pointing arrow). The dark horizontal band at the center of the spectra corresponds to the position of the opaque coronagraphic mask, while the two dark vertical bands around 1.4 and 1.9~\mic are due to the atmospheric absorption. The display scale is identical for both spectra. Strong diffraction residuals are clearly visible on either side of the coronagraphic mask for the LSC, while the apodizer has almost completely suppressed diffraction for the ALSC.}
  \label{fig:example_data}
\end{figure}

\begin{figure}
  \centering
  \includegraphics[width=0.45\textwidth]{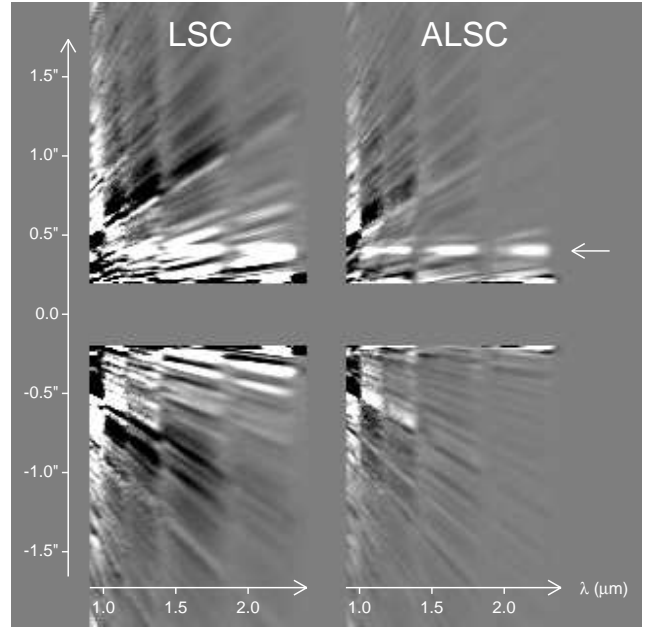}
  \caption{Same simulated data as Fig.~\ref{fig:example_data} after applying the spectral deconvolution data analysis method. The fake companion at a separation of 0.4\as is now clearly visible as a horizontal signal 0.2\as above the coronagraphic mask position (left-pointing arrow). Although the diffraction residuals of the LSC have been attenuated, they are still visible and obviously bias the signal of the companion.}
  \label{fig:example_data_da}
\end{figure}

The outputs of the diffractive simulation are normalized images of the PSF and the coronagraphic PSF that need to be scaled to actual photometric values and combined to produce a spectrum. A photometric code is used to convert the diffractive output into a realistic observation of a star-planet system. The code takes the photometry of the objects being observed, the overall instrumental transmission budget (including the apodizer and the quantum efficiency of the detector), and different noise sources into account. These sources are the read-out noise (10~e$^{-}$/read), the flat field noise (0.3\%, typical accuracy obtained on the Hawaii--2RG detector flat fields of IRDIS), and the photon noise. Thermal background from the instrument ($\sim$340 photon~sec$^{-1}$~pixel$^{-1}$) and sky are also included. The sky is modeled by a constant emission below 2~\mic and a black body at longer wavelengths \citep{lena1998}. The emission values are calculated to match the typical sky brightness in the near-IR provided by ESO\footnote{\url{http://www.eso.org/gen-fac/pubs/astclim/paranal/skybackground/} } ($J$ = 16.5, $H$ = 14.4, $K_{\mathrm{s}}$ = 13.0). The temporal evolution of the OH lines is not considered in these simulations.

\begin{figure*}
  \centering
  \includegraphics[width=1\textwidth]{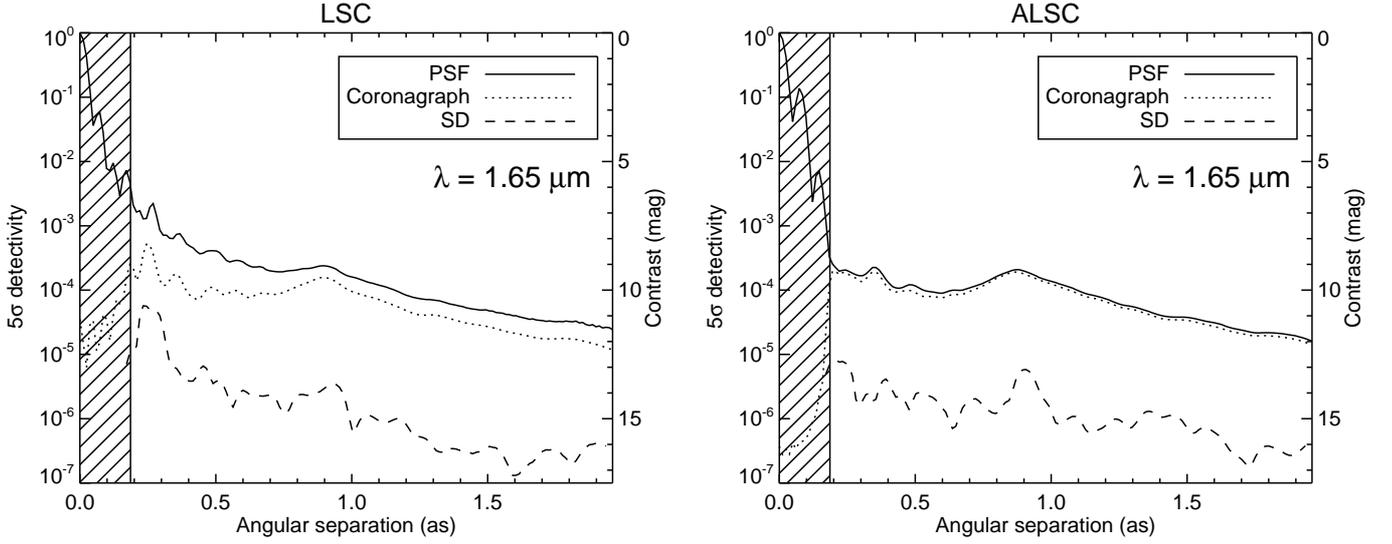}
  \caption{5$\sigma$ detectivity for the LSC (left) and ALSC (right) after attenuation of the speckles with the SD method. The data was simulated assuming bright AO guide star and without detection noise to be in a regime truly limited by speckle noise. The plots show the contrast curves at $\lambda = 1.65$~\mic with the PSF profile (plain), the coronagraphic profile (dotted), and the 5$\sigma$ noise level after SD (dashed). The hatched area corresponds to the location of the opaque coronagraphic mask.}
  \label{fig:contrast_plots}
\end{figure*}

Observations representing one hour of total exposure time are created that combine various stars and planets. The stars are G0, K0 and M0 at 10~pc, simulated using standard Kurucz models \citep{kurucz1979,castelli2003}. The planets cover a wide range of effective temperatures (\teff = 400--2000~K) and surface gravity (\logg = 3.5--5.0) values. They are simulated using the BT-SETTL10 library\footnote{\url{http://phoenix.ens-lyon.fr/Grids/BT-Settl/}} of models \citep{allard2010} at solar metallicity. The planets are introduced at angular separations of 0.3\as, 0.4\as, 0.5\as, 0.75\as, 1.2\as, and 1.5\as, which translates respectively to 3, 4, 5, 7.5, 12, and 15~AU for a star at 10~pc. The simulated observations cover a range of planet/star contrasts from 6.3 to 17.9~mag (averaged $YJHK_{\mathrm{s}}$ contrast). The contrast of all the simulated planets with respect to G0, K0, and M0 stars is plotted in Figure~\ref{fig:contrast_all_planets}. Examples of spectra of a 1400~K companion at 0.4\as from a G0 star at 10~pc are shown in Fig.~\ref{fig:example_data}. 

Finally, two essential calibrations were specifically simulated. The wavelength calibration was taken into account by simulating observations of an extended source (such as an integrated sphere) illuminated by laser lines at 1.05, 1.20, 1.40, 1.68, and 2.10~\mic: a linear fit to the detected positions of the peaks on the detector was used to attribute a wavelength values to each pixel. Observation of an A0 spectroscopic standard was also simulated for correction of the simulated atmospheric absorption features. In that case, no coronagraph was simulated since we wanted to measure the stellar spectrum.

\section{Speckle noise attenuation with spectral deconvolution}
\label{sec:speckle_noise_attenuation_sd}

\begin{figure}
  \centering
  \includegraphics[width=0.5\textwidth]{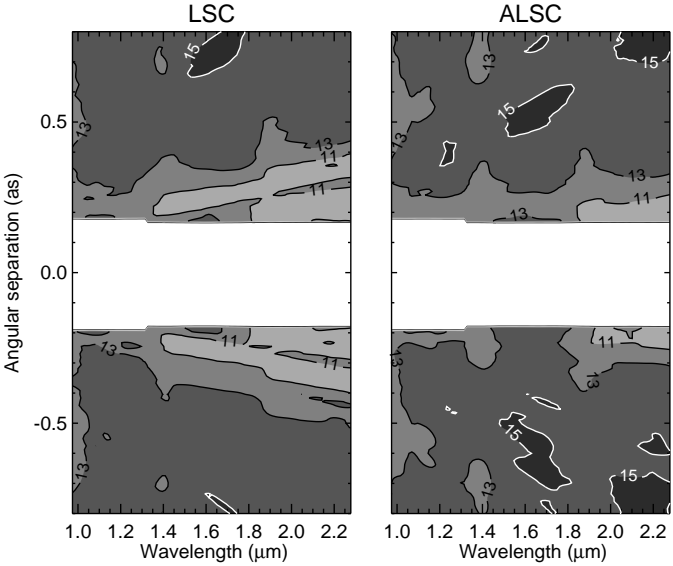}
  \caption{5$\sigma$ noise level after SD at all wavelengths, given in magnitude difference with respect to the star, and averaged for all our simulated data sets. Only the central area between -0.8\as and 0.8\as is plotted since the gain of the ALSC is mainly visible at small angular separation.
}
  \label{fig:contrast_contours}
\end{figure}

When trying to detect faint point sources in the close vicinity of bright stars, observations are limited by the speckle noise, which has a higher amplitude than the signal being sought. The speckles not only limit the ability to detect point sources, but they also bias any flux measurement made on a detected point source. When it comes to characterizing exoplanets with spectroscopy, the bias should be as low as possible to limit errors in the spectral analysis. With an IFS, where both spatial and spectral information is available, a method called the \emph{Spectral Deconvolution} (SD) was proposed and successfully tested on sky \citep{sparks2002,thatte2007}. We presented an adaptation of the SD method for high-contrast LSS with coronagraphy \citep{vigan2008} and showed that it is efficient at suppressing the stellar contribution (speckles, halo), leading to a significant gain in contrast. The method was also successfully tested on the sky with VLT/NaCo LSS data \citep{vigan2012} at moderate contrast and angular separation.

Figure~\ref{fig:example_data} shows simulated data for the LSC and ALSC before applying SD. The LSC spectrum is dominated by the diffraction residuals at close angular separation within the AO-corrected region, creating bright oblique lines that completely hide the signal of the companion simulated at 0.4\as. Thanks to the apodization that suppresses the diffraction, the ALSC spectrum is dominated by the quasi-static speckles that are at a much lower level than the diffraction pattern. In that data the signal of the simulated companion is easier to detect by eye, but it is still biased by the presence of the speckles. Then the SD method is implemented as follows.

\begin{enumerate}
  \item All spectral channels at $\lambda_{i}$ are rescaled by a factor $\alpha_{i} = \lambda_{0}/\lambda_{i}$, where $\lambda_{0}$ is a reference wavelength of 950~nm. The spectrum obtained is referred to as the rescaled spectrum (see Fig.~2 of \citealt{vigan2008} for an illustration).

  \item A model spectrum of the star is created by averaging the rescaled spectrum in all spectral channels.

  \item Then this model spectrum is fitted in amplitude to all spatial channels of the rescaled spectrum. During this process, the signal of the companion is masked in the fit to avoid any bias (see Sect.~4.2.1 of \citealt{vigan2008}). The result is a synthetic spectrum of the star free of the companion signal.

  \item Finally, the synthetic spectrum is subtracted to the rescaled spectrum to remove the star contribution, and the spectral channels are scaled back to their original size to obtain the final spectrum.
\end{enumerate}

\noindent This implementation makes no assumption about the spectrum of the star, the companion, or the atmospheric absorption. Figure~\ref{fig:example_data_da} illustrates the result of the SD on the LSC and ALSC data. While in both spectra the signal of the companion now dominates the residuals, the number of residuals is higher for the LSC than for the ALSC. The diffraction residuals are still particularly strong, and we can qualitatively assume that the spectral extraction of the companion will be more biased from the LSC data than from the ALSC data.

Indeed, one of the drawbacks of the SD method applied to LSS data is that because of the PSF chromaticity, speckles or diffraction residuals that are hidden behind the coronagraphic mask at short wavelengths may appear outside of it at longer wavelengths, creating features at small angular separation that will be extremely difficult to subtract, regardless of the method employed. For this reason, any instrumental device that will decrease the amount of diffraction residuals at close angular separation will bring a significant advantage to the spectral extraction.

To quantify the gain brought by SD for the attenuation of speckle noise, we simulated observations with a bright AO guide star ($V = 4$) and without detection noise. This ensures that we truly are in a regime limited by speckle noise rather than other sources of noise, in particular at separations above 1.0\as. Observations were simulated for our five realizations of quasi-static aberrations to measure the average attenuation. A comparison of the speckle noise attenuation with the LSC and the ALSC is given in Fig~\ref{fig:contrast_plots}. The plots show the level of noise after SD at 1.65~\mic for one of the realizations of quasi-static aberrations with respect to the PSF and the coronagraphic profile. The noise is calculated in a sliding box with a width equal to the width of the slit and of height equal to \lsd. For the LSC, the diffraction residuals are strong on the coronagraphic profile between 0.2\as and 0.6\as. The 5$\sigma$ noise level after SD allows reaching $4 \times 10^{-6}$ (13.5~mag) at 0.5\as, but there is a significant increase for smaller separation due to the poorly subtracted diffraction residuals: at 0.3\as a level of only $4 \times 10^{-5}$ (11.0~mag) can be reached. For the ALSC, the diffraction is completely suppressed and no residuals are visible on the coronagraphic profile. The performance of SD is thus much better at a separation of 0.3\as, where a level of $2 \times 10^{-6}$ (14.2~mag) can be reached, which is 3.2~mag better than the LSC. At larger separations ($\geq 0.5\as$), both designs offer similar performances, with even a 1.5~mag advantage for the LSC between 1.2\as and 1.6\as.

When looking at the full wavelength range close to the star in Fig.~\ref{fig:contrast_contours}, the effect of the diffraction residuals at small separation for the LSC are visible as oblique lines, which corresponds to the position of the Airy diffraction rings that have not been completely subtracted with SD. In $H$- and $K_{\mathrm{s}}$-bands, the SD only allows reaching a contrast of $\sim$11~mag. Thanks to the apodization, the ALSC does not show any sign of diffraction residuals from $Y$- to $H$-band, allowing reaching contrasts of $\sim$13~mag at 0.3\as. In $K_{\mathrm{s}}$-band, the contrast decreases to 11~mag because of the PSF core that is not fully covered by the opaque coronagraphic mask. This is the case because the ALSC design has been optimized for a wavelength of 1.6~\mic, so that above $\sim$1.9~\mic, the mask becomes too small to hide the PSF core where the apodization has concentrated most of the energy. But even in this band, the area covered by the diffraction ring is small (0.1\as) compared to the diffraction residuals of the LSC, which extend over at least 0.3\as.

\section{Spectral extraction}
\label{sec:spectral_extraction}

\begin{figure*}
  \centering
  \includegraphics[width=1\textwidth]{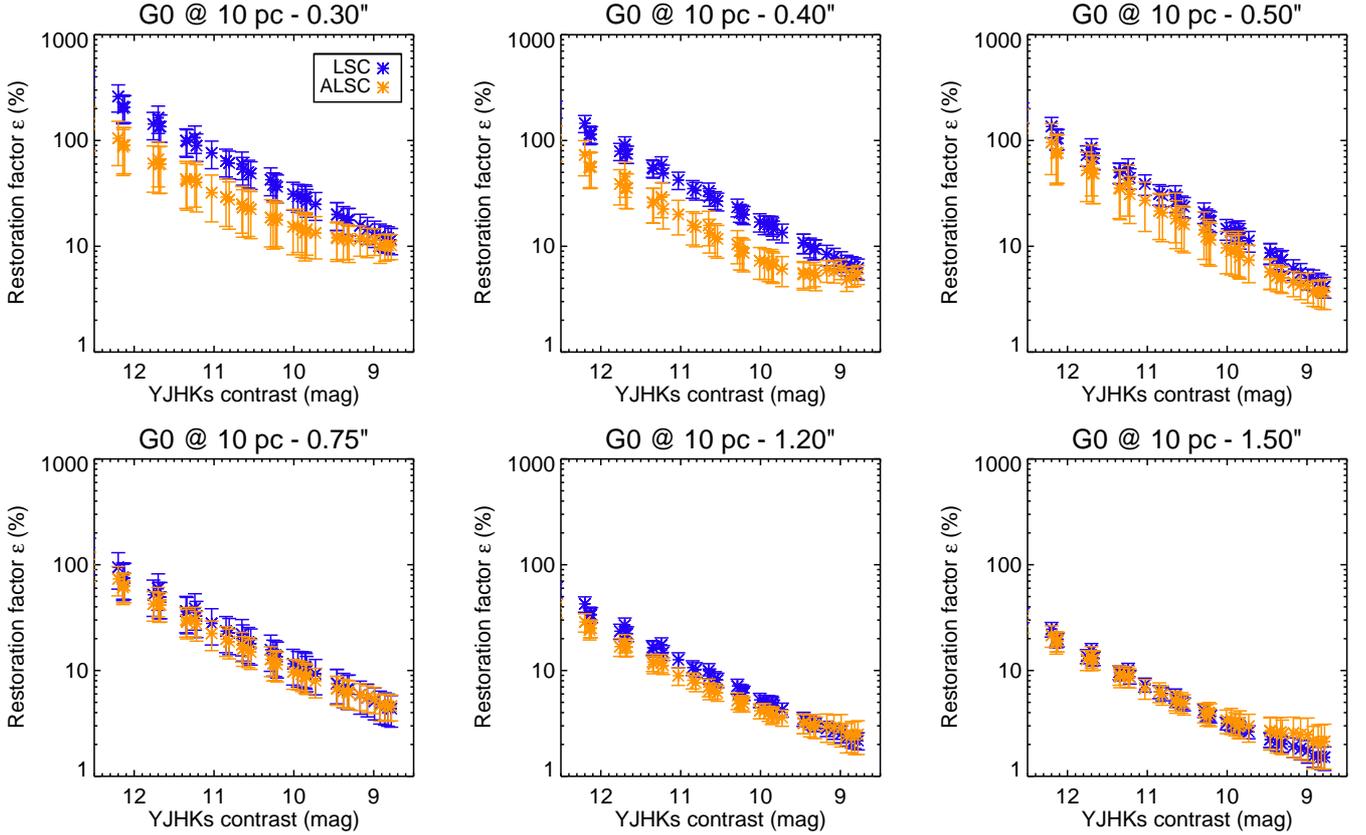}
  \caption{Restoration factor for LSC (blue) and ALSC (orange) as a function of the average $YJHK_{\mathrm{s}}$ contrast of the simulated planets as increasing angular separations from a G0 star at 10~pc. The $1\sigma$ error bars are obtained by introducing the same planets in the 5 independent data sets, at two symmetric positions on either sides of the star (10 measurements in total).}
  \label{fig:restoration_contrast}
\end{figure*}

We have seen in the previous section that thanks to the suppression of the diffraction, the ALSC offers a better contrast performance at a very small angular separation. However, this result does not fully demonstrate that the ALSC is a better concept than the LSC for characterizing exoplanets. Indeed, one of the requirements when doing spectroscopy is to be able to extract a clean spectrum of the planetary companion, i.e., a spectrum that is not biased by the signal of the star contained in the speckles. In this section we demonstrate that the ALSC allows characterizing fainter, thus cooler, objects than the LSC at very small angular separations. In Sect.~\ref{sec:method} we explain the methodology used to quantify the characterization capabilities of the two concepts, in Sect.~\ref{sec:results_spectral_extraction} we detail the results and in Sect.~\ref{sec:limits_teff_logg} we quantify the ability to evaluate the physical parameters \teff and \logg.

\subsection{Method}
\label{sec:method}

As explained in Sect.~\ref{sec:high_contrast_spectroscopy_simulations}, simulations of various planetary systems were produced, covering a wide range of contrast ratios. For each combination of star spectral type, planet \teff, \logg, and angular separation, a fake spectrum is produced using each of the five simulated data sets. Spectral deconvolution is then applied to each spectrum to remove the stellar halo and speckles, and the spectrum of the planet is extracted. For the actual extraction, the flux in each spectral channel is integrated in a \lsd-wide aperture centered on the position of the planet. The spectrum of the spectroscopic standard was extracted with the same procedure and divided by a black body spectrum at the appropriate \teff.

Once the planet spectrum is extracted, different means are available to measure the quality of the extraction. Since our work is within the realm of numerical simulations, the input spectrum is perfectly known and can be used as the reference. For this comparison, we follow \citet{pueyo2011} and define a restoration factor to quantify the relative spectro-photometric error:

\begin{equation}
  \label{eq:relative_error_sed}
  \epsilon = \sqrt{ N_{\lambda} \sum_{p=1}^{N_{\lambda}} \left(s_{p}^{\mathrm{atm}}\right)^2\left( \frac{I_{p}}{\sum_{k=1}^{N_{\lambda}} s_{k}^{\mathrm{atm}}I_{k}^{\mathrm{ref}}} -  \frac{I_{p}^{\mathrm{ref}}}{\sum_{k=1}^{N_{\lambda}} s_{k}^{\mathrm{atm}}I_{k}^{\mathrm{ref}}} \right)^2 },
\end{equation}

\noindent where $I^{\mathrm{ref}}$ is the input spectrum, $I$ the spectrum extracted after spectral deconvolution, $s^{\mathrm{atm}}$ the measure of the atmospheric absorption derived from the spectroscopic standard, and $N_{\lambda}$ the number of spectral channels. This factor measures how well the overall shape of the spectrum is recovered, up to a scaling constant. A low value for $\epsilon$ denotes a better restoration. The lower S/N obtained at the wavelengths with strong atmospheric absorption is taken into account by weighting the measurements with $s^{\mathrm{atm}}$ at each wavelength.

We note that the spectral deconvolution is not able to perfectly estimate and subtract the speckles in LSS data, leaving some residuals that are clearly visible in the example of Fig.~\ref{fig:example_data_da}. Between 0.95 and 1.0~\mic, the speckles subtraction appears significantly less efficient than at longer wavelengths, inducing a large flux depletion in the extracted spectrum. Some flux depletion also appears at 2.3~\mic close to the edge of the coronagraph mask, at the position where some bright Airy rings are located. The location of these poorly subtracted speckle residuals suggest a possible edge effect induced by the finite number of wavelengths simulated to create the spectra. Given the large amount of time required to simulate LSS data, we could not investigate wether this is a purely numerical effect or if there is actually a limitation for the SD at the lower end, and less significantly at the higher end, of the spectra. A careful analysis of real, extreme AO, high-contrast LSS data will certainly provide an answer when such data becomes available. To avoid any artificial bias in our simulations, the data points below 1.0~\mic were not considered when calculating the restoration factor. Results obtained when including these data points are almost identical to the ones presented above, except for notably higher detection limits in \teff at 0.5\as. This separation corresponds to the position of the AO cutoff at 0.95~\mic, where the bias is the strongest.

A more physical alternative to evaluate the performance of both concepts is to compare the extracted spectrum to synthetic libraries of atmosphere models. Spectra are compared to the libraries AMES-DUSTY \citep{allard2001}, AMES-COND \citep{allard2003}, and BT-SETTL10 \citep{allard2010} with \teff from 300 to 2500~K and \logg from 2.5 to 6.0 to find the best match using a $\chi^2$ minimization. Before calculation of the $\chi^2$, the models from the libraries are convolved with a Gaussian kernel to smooth their resolution down to that of the simulated spectrum, and interpolated on the same wavelength grid. Then they are compared to the extracted spectrum corrected for the atmospheric absorption. To avoid any bias in the fit from areas of strong atmospheric absorption, the regions of the spectrum where the atmospheric transmission is lower than 80\% are given a weight of zero when calculating the $\chi^2$. Since our simulated spectra are not flux-calibrated -- i.e., we infer \teff and \logg only from the overall shape of the spectrum -- we add a normalization constant than can vary freely to minimize the discrepancy between the spectrum and the model. The final estimation of \teff and \logg are that of the best-match model in the libraries.

Finally, we note that although the AO cutoff occurs at a separation of 20\lsd, which translates into $\sim$0.5\as and $\sim$1.2\as respectively at 0.95~\mic and 2.3~\mic, we do not exclude or treat the data located outside of the AO control radius differently. The use of spectral deconvolution is not limited to the AO-corrected region, since even outside of this region the halo of the star needs to be attenuated for a good spectral extraction. It means that the extracted companion spectrum is taken as a whole, including parts of the data that are not within the AO control radius. Similarly, when comparing the extracted spectrum to synthetic libraries of models, we do not exclude any of the data from the fit based on its position with respect to the AO control radius. Although it could be expected that fitting only data within this region might yield a better estimation of \teff and \logg, \citet{patience2012} have shown on VLT/SINFONI data of a dozen substellar companions that single-band fits provide large systematic errors compared to multiband fits of the same spectrum. This result is relevant whether extreme AO is used (as in IRDIS) or not (as in SINFONI) because the biases in the estimation of the physical parameters are induced by the single-band analysis, which necessarily provides less information than a full near-IR multiband analysis. To avoid such biases, especially considering that we use very low-resolution spectra, we perform a multiband fit for the estimation of the parameters of our simulated objects.

\subsection{Results for spectral extraction}
\label{sec:results_spectral_extraction}

\begin{figure}
  \centering
  \includegraphics[width=0.5\textwidth]{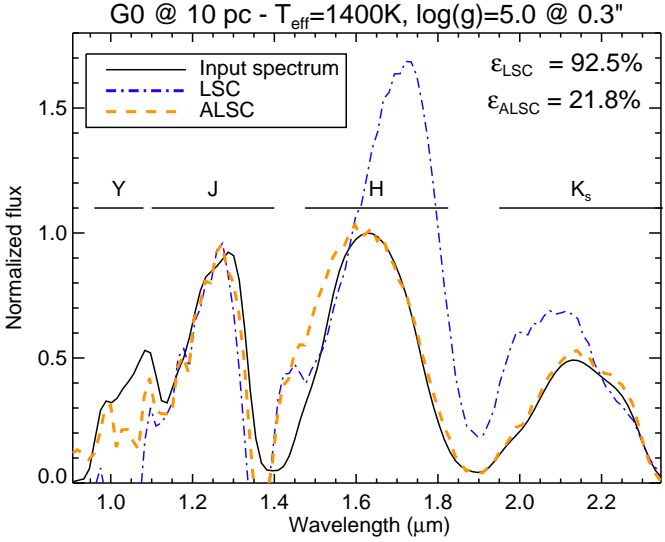}
  \caption{Spectral extraction of a companion with \teff~=~1400~K and \logg~=~5.0 simulated at 0.3\as from a G0 star at 10~pc (contrast of 10.6~mag) using the LSC and ALSC coronagraphs. The plot shows input planet spectrum (black, plain), and the extracted planet spectrum from data simulated with the LSC (blue, dash-dotted) and the ALSC (orange, dashed). The restoration factor $\epsilon$ has a lower value with the ALSC, denoting a better restoration of the planet spectrum.}
  \label{fig:restoration_example}
\end{figure}

\begin{figure}
  \centering
  \includegraphics[width=0.5\textwidth]{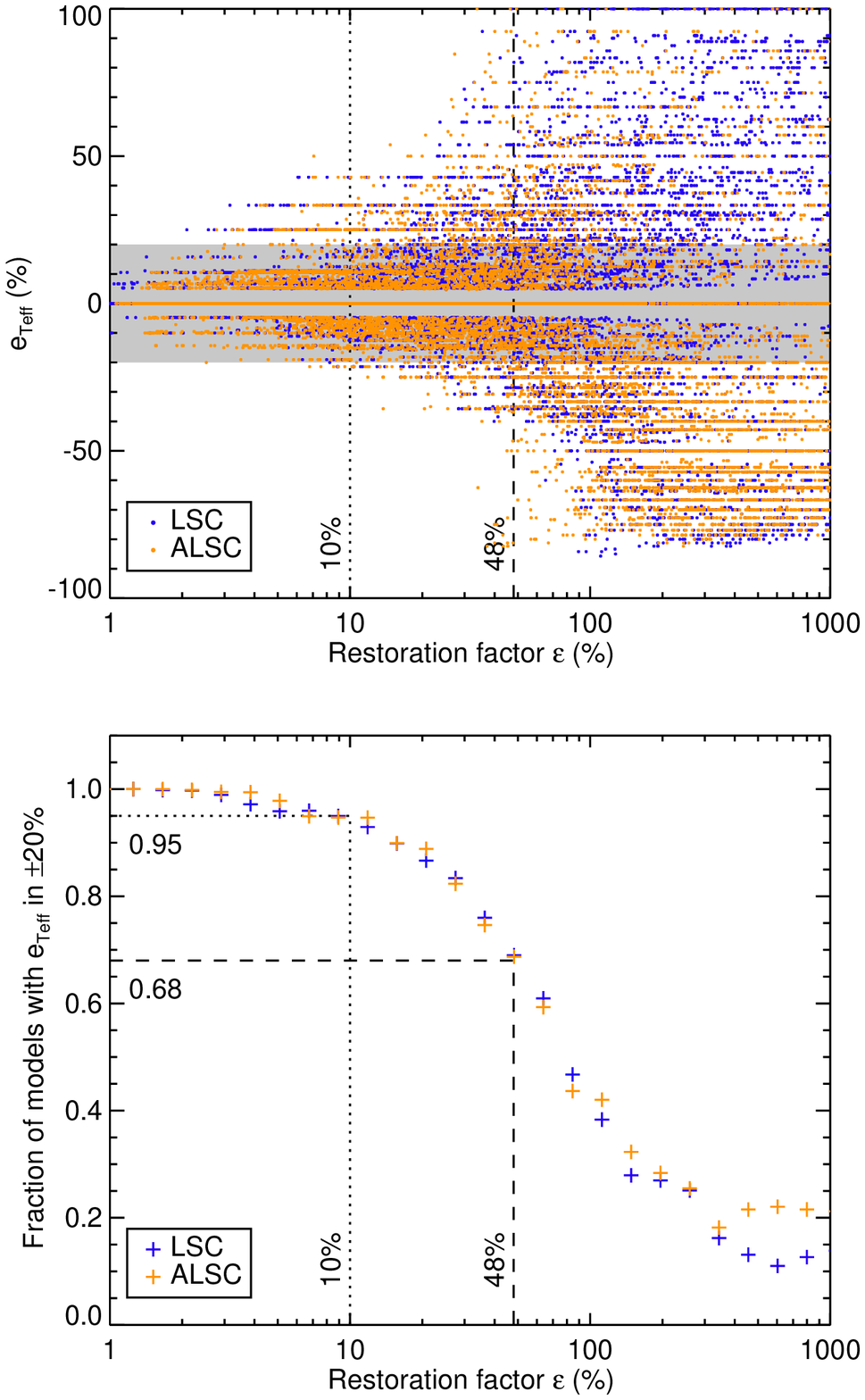}
  \caption{Relative error on the determination of \teff ($e_{T_{e\!f\!f}}$) as a function of the restoration factor (top) for all simulated planets, stars, and separations, and both coronagraphs (LSC in blue, ALSC in orange). The gray area corresponds to a $\pm$20\% interval. The bottom plot gives the fraction of planets for which $e_{T_{e\!f\!f}}$ is within $\pm$20\%, as a function of the restoration factor. The restoration factors below which 68\% and 95\% of the models have $e_{T_{e\!f\!f}}$ within $\pm$20\% on \teff are 48\% and 10\%, respectively.}
  \label{fig:tefferr_restoration}
\end{figure}

The ability to extract a clean spectrum depends on the level of the planet signal with respect to the level of the speckle noise. The driving parameters are the overall contrast of the planet with respect to the star and the angular separation. Indeed, for a given contrast, the spectrum of a planet at a very small angular separation will be more difficult to extract than for a planet at several arcseconds. Figure~\ref{fig:restoration_contrast} shows the value of the restoration factor as a function of the average contrast difference between a G0 star at 10~pc and all the simulated planets in the data. The $1\sigma$ error bars are obtained by introducing each fake planet in the five independent data sets, at two symmetric positions on either side of the star (e.g. +0.4\as and -0.4\as, see Fig.~\ref{fig:example_data}), resulting in ten measurements for each planet. 

As explained in the previous section, a smaller restoration factor denotes an extracted spectrum of better quality. From Fig.~\ref{fig:restoration_contrast} it clearly appears that the ALSC offers a significantly better restoration of the spectra at small angular separation. The effect is the most significant at 0.3\as, and slowly decreases up to 0.75\as where the restoration factors have identical values for both coronagraphs. The cutoff of the extreme AO system occurs at $\sim$0.5\as and $\sim$1.2\as at 0.95~\mic and 2.3~\mic, respectively, so in any case the differences between the LSC and ALSC coronagraph should be expected below these separations. With the LSC, the data at small separation is mostly limited by the diffraction residuals (i.e. the Airy rings) that are at a higher level than the quasi-static aberrations, inducing a large bias in the spectral extraction, thus resulting in a high restoration factor. Since the ALSC suppresses diffraction, the extraction of the planet spectrum becomes mostly limited by the quasi-static speckles, for which the level is lower than the diffraction residuals but depends on the instrumental aberrations introduced in the data. For our five different data sets, the amount of aberrations introduced is identical, but their random seed is different, creating variations between the data sets.

Finally, Fig.~\ref{fig:restoration_example} illustrates the difference in restoration between the LSC and ALSC for a specific case: a companion with \teff~=~1400~K and \logg~=~5.0 is simulated at 0.3\as from a G0 star at 10~pc (contrast of 10.6~mag) using each coronagraph. For both coronagraphs, the quasi-static aberrations are identical. When compared to the input planet spectrum, the spectrum extracted after SD is restored much better with the ALSC ($\epsilon = 21.8\%$) than with the LSC ($\epsilon = 92.5\%$). The LSC spectrum is clearly biased in $Y$-, $H$-, and $K_{\mathrm{s}}$-bands, with too much flux being subtracted or left over, while the ALSC spectrum does not show any major spurious features.

\subsection{Limits on \teff and \logg}
\label{sec:limits_teff_logg}

\begin{figure*}
  \centering
  \includegraphics[width=1.0\textwidth]{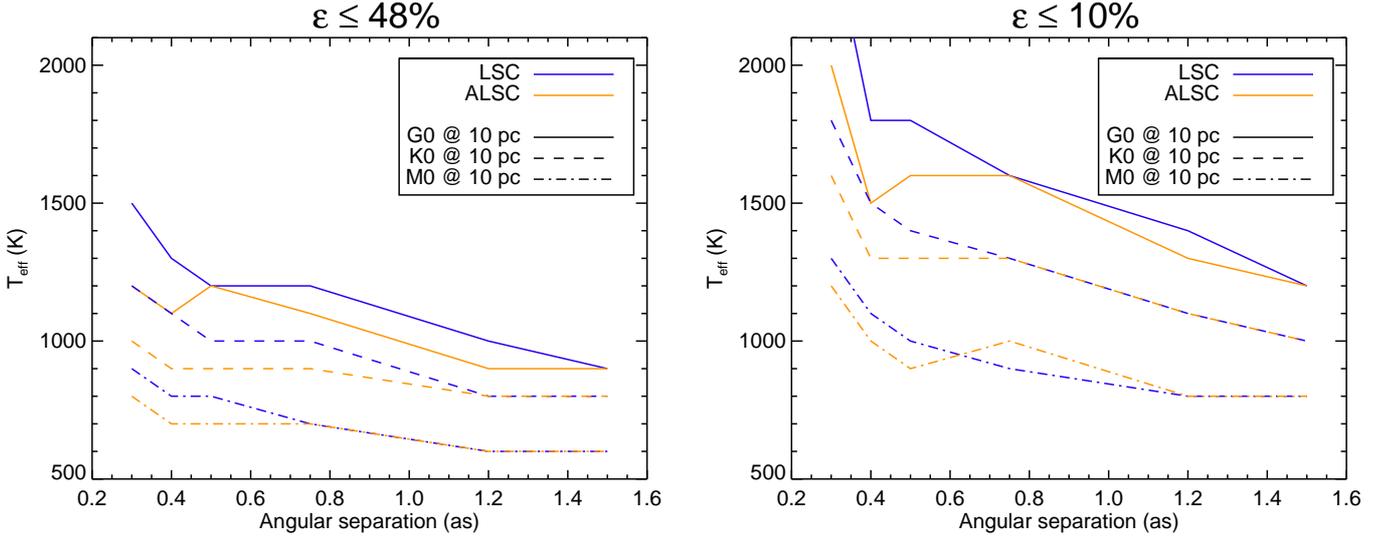}
  \caption{Lower limits in \teff with the LSC (blue) and ALSC (orange) for all models with $\epsilon \le 48\%$ and $\epsilon \le 10\%$ around G0 (plain), K0 (dashed), and M0 (dash-dotted) stars at 10~pc as a function of angular separation. The cutoffs on $\epsilon$ correspond to the limit where more than 68\% and 95\% of the simulated models have $e_{T_{e\!f\!f}}$ within $\pm$20\%. For $\epsilon \le 10\%$, the LSC curve at 0.3\as reaches 2500~K.}
  \label{fig:teff_limit}
\end{figure*}

\begin{figure*}
  \centering
  \includegraphics[width=1.0\textwidth]{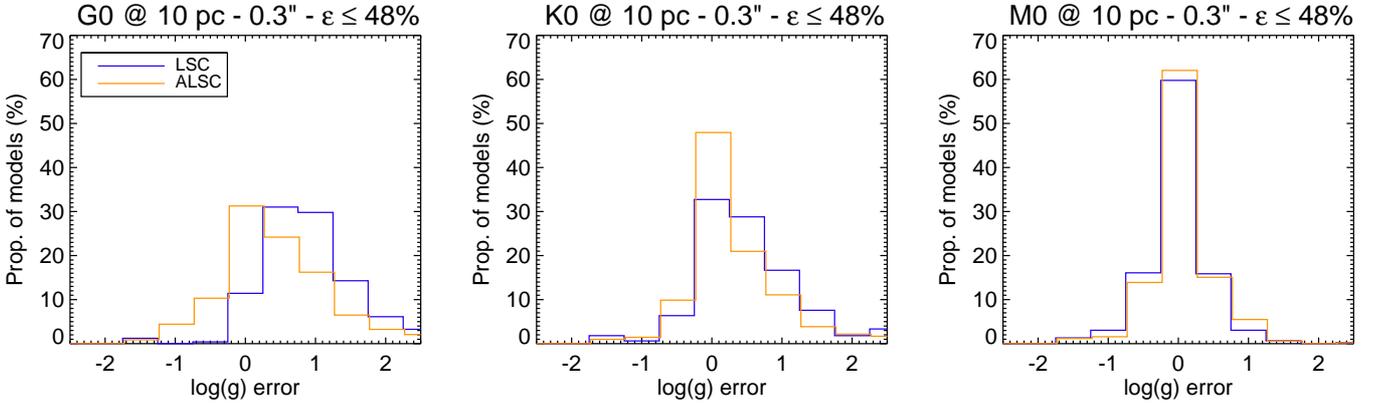}
  \caption{Distribution of the errors on the determination of \logg with the LSC (blue) and ALSC (orange) for all models with $\epsilon \le 48\%$ at 0.3\as around G0 (left), K0 (center) and M0 (right) stars at 10~pc.}
  \label{fig:logg_limit}
\end{figure*}

\begin{figure*}
  \centering
  \includegraphics[width=1.0\textwidth]{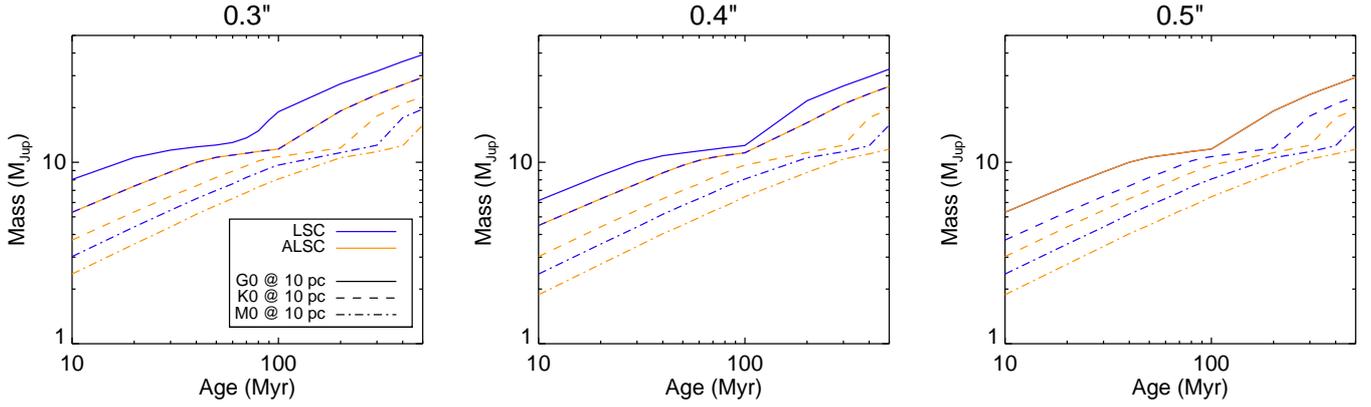}
  \caption{Mass of the planets that could be characterized with the LSC (blue) and ALSC (orange) at 0.3\as, 0.4\as, and 0.5\as from G0 (plain), K0 (dashed), and M0 (dash-dotted) stars at 10~pc, assuming $\epsilon \le 48\%$ ($e_{T_{e\!f\!f}}$ within $\pm$20\% for 68\% of the models). The mass is obtained using AMES-COND evolutionary models \citep{baraffe2003}. The curves for the LSC and the ALSC above 0.5\as are almost identical.}
  \label{fig:mass_limit}
\end{figure*}

To compare the LSC and ALSC in terms of potential scientific output, it is interesting to study the precision with which the physical parameters of a planet can be determined, i.e. its \teff and \logg in our simulations. For this purpose we link the restoration factor to the precision with which the planets \teff can be determined. The goal is to determine a threshold on the restoration factor below which the spectrum can be considered as usable for science, i.e., for determining the physical parameters of the planet from a low-resolution spectrum.

\subsubsection{Threshold}
\label{sec:threshold}

Figure~\ref{fig:tefferr_restoration} (top) shows the relative error on the determination of \teff as a function of the restoration factor for all simulated planets, stars, and angular separations. The restoration factor and \teff were estimated using the procedure described in Sect.~\ref{sec:method}. When the restoration factor increases, the relative error on \teff shows a clear increase going from $\pm$20\% to $\pm$100\% or more. The cutoff at $-80\%$ is the result of the limited grid of models in \teff (300--2500~K) to which we compare the extracted spectrum: the largest error occurs when a simulated planet at 2000~K is best fitted by a model at 300~K, resulting in a relative error of $-85\%$. This extreme case will of course occur only at high contrast where the extracted spectrum is dominated by the residual speckles and the $\chi^2$ minimization is unreliable. Indeed, this lower cutoff is reached for restoration factors above 100\%, which correspond to spectra that are extremely biased as can be seen on Fig.~\ref{fig:restoration_example}. When the \teff is not determined correctly (null error), the grid step of 100~K induces a minimal relative error of 4\% that is visible on Fig.~\ref{fig:tefferr_restoration} (top) with an empty range of $\pm$4\% around zero.

We chose to consider a threshold of $\pm$20\% for the relative error on \teff ($e_{T_{e\!f\!f}}$), corresponding to an error of 80~K for the coldest simulated planet (400~K), which is approximately equal to the grid step. Figure~\ref{fig:tefferr_restoration} (bottom) shows the fraction of models for which $e_{T_{e\!f\!f}}$ is within $\pm$20\% as a function of the restoration factor. Both coronagraphs show a similar variation: the fraction of models with $e_{T_{e\!f\!f}}$ within $\pm$20\% is high for restoration factors below $\sim$10, and then it steeply declines in the range 10--100\% to reach a fraction of 0.3 at 200. We use this plot to determine the value of the restoration factor below which a given fraction of the simulated models has $e_{T_{e\!f\!f}}$ within $\pm$20\%. We want to establish the range of parameters where most of the models can be characterized, so we select fractions of 0.68 (equivalent to 1$\sigma$ for a Gaussian noise distribution) and 0.95 (2$\sigma$), which respectively corresponds to restoration factors $\epsilon = 48\%$ and $10\%$. For example, this means that for a simulated spectrum with $\epsilon \le 48\%$, we have a 68\% chance to determine its \teff at a precision better than 20\%.

As noted above, the distributions of $e_{T_{e\!f\!f}}$ as a function of $\epsilon$ in Fig.~\ref{fig:tefferr_restoration} is independent of the type of coronagraph when plotting all models together. The way the restoration factor is designed should not allow the value to depend on outside parameters, such as the star spectral type, the coronagraph, or the angular separation. However, when separately considering the simulations for different stars and different separations, we observed some variations between the LSC and ALSC on the thresholds needed to obtain a large number of models with $e_{T_{e\!f\!f}} \le 20\%$. But these variations are artificialy biased by the choice of the star and the angular separation. Considering a given star and angular separation in Fig.~\ref{fig:tefferr_restoration} is equivalent to setting the level of signal from the speckles and stellar halo. As we have seen in Sect.~\ref{sec:results_spectral_extraction}, the ALSC generally provides lower values of $\epsilon$ for given values of the separation and contrast ratio, so when calculating the fraction of models with $e_{T_{e\!f\!f}} \le 20\%$ in a given bin of $\epsilon$ values, the number of measurements in that bin will be different for the LSC and ALSC (more points at lower $\epsilon$ values for the ALSC), ultimately biasing the calculation of the thresholds on $\epsilon$. To circumvent this problem, the thresholds on $\epsilon$ have been calculated by combining all models, stars, and separations, resulting in a large number of measurements ($>100$) in each bin and for each type of coronagraph.

\subsubsection{Estimation of \teff and \logg}
\label{sec:estimation_teff_logg}

In Fig.~\ref{fig:teff_limit} we plot the lowest \teff for all models with $\epsilon \le 48\%$ and $\epsilon \le 10\%$ around G0, K0, and M0 stars at 10~pc, as a function of angular separation. Models with restoration factors above the cutoffs are colder (fainter planets), but their $e_{T_{e\!f\!f}}$ is larger than 20\% in absolute value, so they are not considered. The difference between the LSC and ALSC at small angular separation that was noted in Sect.~\ref{sec:results_spectral_extraction} is clearly visible. At 0.3\as the ALSC allows characterizing models that are at least 100~K colder than the LSC, while at larger angular separations, the difference between the two coronagraphs decreases or disappears. In both cases, the largest difference occurs at higher contrast (brighter stars), which is expected considering that for a bright star, the diffraction residuals in the LSC will produce higher contamination and will bias the \teff estimation more strongly.

The second important parameter that defines the properties of a given model is \logg. In Fig.~\ref{fig:logg_limit} we represent the distribution of the errors on the determination of \logg, assuming $\epsilon \le 48\%$, for models simulated at 0.3\as from G0, K0, and M0 stars at 10~pc. In the case of a G0 at 10~pc, the distributions are broad for both coronagraphs, which means that the determination of \logg is not accurate, but the ALSC reaches a maximum for a null error on \logg, while the LSC is biased toward positive errors on \logg. As the contrast ratio decreases, the distributions for both coronagraphs become more peaked around zero. For more favorable contrast ratios, e.g. for the M0 star at 10~pc, the LSC and ALSC are equivalent for determining \logg.

We note that for determining \teff and \logg in low-contrast conditions, i.e. bright planets at large angular separation, errors of up to 20\% can occur even for very low values of the restoration factor (Fig.~\ref{fig:tefferr_restoration}). This means that even for high-fidelity spectra extracted from the data, the determination of \teff can be uncertain using these very low-resolution spectra. A possible improvement would be to consider not only the shape of the spectrum for determining of the parameters, but also its overall luminosity and contrast with respect to the star. An improved $\chi^2$ taking the flux ratio between the star and the models from the library into account would give more weight to the models with a contrast closer to the spectrum extracted from the data, and possibly limit the errors at low contrast. To remain as general as possible, we decided to keep the comparison of the extracted spectra as simple as possible for this work, using only the overall shape.

Finally, to have a clearer idea of the direct scientific output, the detection limits in \teff can be converted into a planet mass using evolutionary models. Figure~\ref{fig:mass_limit} gives the planet mass as a function of age at three angular separations, obtained from the \teff limits with $\epsilon \le 48\%$ from Fig.~\ref{fig:teff_limit} and the AMES-COND evolutionary models \citep{baraffe2003}. At 0.3\as around a 10~Myr-old G0 star at 10~pc, the ALSC reaches a mass of $\sim$5~\MJup, compared to $\sim$8~\MJup for the LSC. Over the whole range of plotted ages and for the different stars, a factor 1.2--1.8 is <maintained in favor of the ALSC at 0.3\as. At larger separation the advantage of the ALSC mostly wears off above 0.5\as.

\section{Conclusion}
\label{sec:conclusion}

High-contrast LSS of faint planetary-mass companions to main-sequence stars is strongly limited by the quasi-static speckle pattern and the diffraction residuals when using a classical Lyot coronagraph. The concept of ALSC that we propose is a significant improvement over the previous LSC coronagraph, at the expense of the decreased throughput (37\%) induced by the use of a pupil apodization. The almost complete suppression of the diffraction enables looking at closer separation and deeper contrast. Indeed, the presence of a slit inside the coronagraphic plane induces a complex energy distribution in the Lyot pupil plane, thereby preventing the use of a simple Lyot stop. In terms of astrophrysical output, the ALSC allows detecting much fainter, i.e., much cooler planets, which necessarily corresponds to lower masses. Given the various uncertainties on the properties of massive planets in wide orbits, any instrumental means improving their spectral characterization is important.

To demonstrate the advantages of the ALSC, the design was optimized for SPHERE/IRDIS, the upcoming differential spectro-imager for the VLT. In this configuration, using a 0.18\as coronagraphic mask with no aberrations, the ALSC provides a gain of a factor $\sim$40 over the LSC at an angular separation of 0.25\as. In the presence of realistic amounts of aberrations ($\sim$50~nm~RMS), the raw coronagraphic contrasts reached with both designs are equivalent, but the absence of diffraction rings for the ALSC is a significant advantage for the subsequent subtraction of the speckles. For a more practical demonstration, we introduced the ALSC into detailed end-to-end simulations of the low-resolution LSS mode of SPHERE/IRDIS. Simulations were performed to obtain data representing 1~h of integration time on different stars at 10~pc with simulated planetary companions at angular separations between 0.3\as and 1.5\as. 

The spectral deconvolution data analysis method was applied to the simulated data to subtract the stellar halo and speckles before extracting the companions spectra. Although generally efficient at estimating and subtracting the diffraction residuals, the SD efficiency is poorer at small angular separation where bright speckles or diffraction residuals might appear from behind the coronagraphic mask, creating features that are difficult to estimate. This is why any device that will decrease the amount of residuals at small angular separation will be an advantage for the spectral extraction. At small angular separations, we showed that the amount of diffraction residuals is lower after SD with the ALSC than with the LSC. At 1.65~\mic, the 5$\sigma$ noise level with respect to the PSF peak at 0.3\as reaches $4 \times 10^{-5}$ (11.0~mag) with the LSC and $2 \times 10^{-6}$ (14.2~mag) with the ALSC. However, at separations larger than 0.50--0.75\as both coronographs offer similar performances. Over the whole $Y$ to $K_{\mathrm{s}}$ range, we showed that ALSC allows reaching contrasts up to $\sim$3~mag deeper at separations smaller than 0.5\as, even in $K_{\mathrm{s}}$-band where some diffraction residuals become significant because of the chromaticity of the PSF.

The improved sensitivity at small angular separation has a direct impact on the spectral extraction of faint companions spectra. A large number of simulated companions with contrasts ranging from 6.3 to 17.9~mag were introduce into the data, and using a restoration factor, which compares quantitatively the input and output companions spectra, we showed that below 0.5\as, the spectra extracted from the ALSC data are systematically of better quality than the ones extracted from the LSC data. The output spectra were all compared to a library of models to estimate their values of \teff and \logg. The comparison with the input values allowed estimating the error on the determination of these parameters and establishing detection limits in terms of \teff around M0, K0, and G0 stars at 10~pc. The results show that at 0.3\as, the limit with the ALSC is systematically at least 100~K lower, inducing an increase in sensitivity to lower mass objects at young ages. The improved limits translate into a gain of a factor 1.2--1.8 on the mass of the objects that could be characterized, which is significant in a range of masses where very few planetary companions have already been imaged.

Although significant improvements over the LSC have been obtained with the ALSC using gray apodization, it is worth mentioning the possibility of using a colored apodization \citep{ndiaye2012} to improve the performance in very broad band. We have seen that one of the limitations in the present work is the chromaticity of the PSF: in $K_{\mathrm{s}}$-band, the opaque coronagraphic mask becomes too small to fully cover the apodized PSF peak, creating residuals in 0.2--0.3\as. The use of an colored apodization could be explored to decrease the impact of the PSF chromaticity by better constraining the size of the core at longer wavelengths.

\begin{acknowledgements}
A. V. acknowledges support from a Science and Technology Facilities Council (STFC) grant (ST/H002707/1). This work is partially based upon work supported by the National Aeronautics and Space Administration under Grant NNX12AG05G issued through the Astrophysics Research and Analysis (APRA) program. M. N. would like to thank R\'emi Soummer and Laurent Pueyo for their support. SPHERE is an instrument designed and built by a consortium consisting of IPAG, MPIA, LAM, LESIA, Laboratoire Fizeau, INAF, Observatoire de Gen\`eve, ETH, NOVA, ONERA, and ASTRON in collaboration with ESO. 
\end{acknowledgements}

\bibliographystyle{aa}
\bibliography{paper}

\begin{thebibliography}{42}
\expandafter\ifx\csname natexlab\endcsname\relax\def\natexlab#1{#1}\fi

\bibitem[{{Aime}(2005)}]{aime2005b}
{Aime}, C. 2005, \aap, 434, 785

\bibitem[{{Allard} {et~al.}(2003){Allard}, {Guillot}, {Ludwig}, {Hauschildt},
  {Schweitzer}, {Alexander}, \& {Ferguson}}]{allard2003}
{Allard}, F., {Guillot}, T., {Ludwig}, H.-G., {et~al.} 2003, in IAU Symposium,
  Vol. 211, Brown Dwarfs, ed. E.~{Mart{\'{\i}}n}, 325--+

\bibitem[{{Allard} {et~al.}(2001){Allard}, {Hauschildt}, {Alexander},
  {Tamanai}, \& {Schweitzer}}]{allard2001}
{Allard}, F., {Hauschildt}, P.~H., {Alexander}, D.~R., {Tamanai}, A., \&
  {Schweitzer}, A. 2001, \apj, 556, 357

\bibitem[{{Allard} {et~al.}(2010){Allard}, {Homeier}, \&
  {Freytag}}]{allard2010}
{Allard}, F., {Homeier}, D., \& {Freytag}, B. 2010, ArXiv e-prints

\bibitem[{{Angel}(1994)}]{angel1994}
{Angel}, J.~R.~P. 1994, \nat, 368, 203

\bibitem[{{Antichi} {et~al.}(2009){Antichi}, {Dohlen}, {Gratton}, {Mesa},
  {Claudi}, {Giro}, {Boccaletti}, {Mouillet}, {Puget}, \&
  {Beuzit}}]{antichi2009}
{Antichi}, J., {Dohlen}, K., {Gratton}, R.~G., {et~al.} 2009, \apj, 695, 1042

\bibitem[{{Baraffe} {et~al.}(2003){Baraffe}, {Chabrier}, {Barman}, {Allard}, \&
  {Hauschildt}}]{baraffe2003}
{Baraffe}, I., {Chabrier}, G., {Barman}, T.~S., {Allard}, F., \& {Hauschildt},
  P.~H. 2003, \aap, 402, 701

\bibitem[{{Beuzit} {et~al.}(2008){Beuzit}, {Feldt}, {Dohlen}, {Mouillet},
  {Puget}, \& {Wildi}}]{beuzit2008}
{Beuzit}, J.-L., {Feldt}, M., {Dohlen}, K., {et~al.} 2008, in {SPIE Conference
  Series}, Vol. 7014

\bibitem[{{Boccaletti} {et~al.}(2008){Boccaletti}, {Carbillet}, {Fusco},
  {Mouillet}, {Langlois}, {Moutou}, \& {Dohlen}}]{boccaletti2008}
{Boccaletti}, A., {Carbillet}, M., {Fusco}, T., {et~al.} 2008, in SPIE
  Conference Series, Vol. 7015

\bibitem[{{Carbillet} {et~al.}(2011){Carbillet}, {Bendjoya}, {Abe}, {Guerri},
  {Boccaletti}, {Daban}, {Dohlen}, {Ferrari}, {Robbe-Dubois}, {Douet}, \&
  {Vakili}}]{carbillet2011}
{Carbillet}, M., {Bendjoya}, P., {Abe}, L., {et~al.} 2011, Experimental
  Astronomy, 30, 39

\bibitem[{{Carbillet} {et~al.}(2008){Carbillet}, {Boccaletti}, {Thalmann},
  {Fusco}, {Vigan}, {Smith}, {Mouillet}, {Dohlen}, {Bendjoya}, \&
  {Ferrari}}]{carbillet2008}
{Carbillet}, M., {Boccaletti}, A., {Thalmann}, C., {et~al.} 2008, in SPIE
  Conference Series, Vol. 7015

\bibitem[{{Carbillet} {et~al.}(2004){Carbillet}, {Verinaud}, {Guarracino},
  {Fini}, {Lardiere}, {Le Roux}, {Puglisi}, {Femenia}, {Riccardi}, {Anconelli},
  {Correia}, {Bertero}, \& {Boccacci}}]{carbillet2004}
{Carbillet}, M., {Verinaud}, C., {Guarracino}, M., {et~al.} 2004, in SPIE
  Conference Series, ed. D.~{Bonaccini Calia}, B.~L. {Ellerbroek}, \&
  R.~{Ragazzoni}, Vol. 5490, 637--648

\bibitem[{{Castelli} \& {Kurucz}(2003)}]{castelli2003}
{Castelli}, F. \& {Kurucz}, R.~L. 2003, in IAU Symposium, Vol. 210, Modelling
  of Stellar Atmospheres, ed. N.~{Piskunov}, W.~W. {Weiss}, \& D.~F. {Gray},
  20P--+

\bibitem[{{Chauvin} {et~al.}(2010){Chauvin}, {Lagrange}, {Bonavita},
  {Zuckerman}, {Dumas}, {Bessell}, {Beuzit}, {Bonnefoy}, {Desidera}, {Farihi},
  {Lowrance}, {Mouillet}, \& {Song}}]{chauvin2010}
{Chauvin}, G., {Lagrange}, A.-M., {Bonavita}, M., {et~al.} 2010, \aap, 509, A52

\bibitem[{{Dohlen} {et~al.}(2008){Dohlen}, {Langlois}, {Saisse}, {Hill},
  {Origne}, {Jacquet}, {Fabron}, {Blanc}, {Llored}, {Carle}, {Moutou}, {Vigan},
  {Boccaletti}, {Carbillet}, {Mouillet}, \& {Beuzit}}]{dohlen2008}
{Dohlen}, K., {Langlois}, M., {Saisse}, M., {et~al.} 2008, in {Ground-based and
  Airborne Instrumentation for Astronomy II}, Vol. 7014, {SPIE Conference
  Series}, 70143L

\bibitem[{{Fusco} {et~al.}(2006){Fusco}, {Petit}, \& {Rousset}}]{fusco2006}
{Fusco}, T., {Petit}, C., \& {Rousset}, G. e.~a. 2006, in SPIE Conference
  Series, Vol. 6272

\bibitem[{{Gonsalves} \& {Nisenson}(2003)}]{gonsalves2003}
{Gonsalves}, R. \& {Nisenson}, P. 2003, \pasp, 115, 706

\bibitem[{{Guyon} {et~al.}(2006){Guyon}, {Pluzhnik}, {Kuchner}, {Collins}, \&
  {Ridgway}}]{guyon2006}
{Guyon}, O., {Pluzhnik}, E.~A., {Kuchner}, M.~J., {Collins}, B., \& {Ridgway},
  S.~T. 2006, \apjs, 167, 81

\bibitem[{{Joos} {et~al.}(2011){Joos}, {Schmid}, {Gisler}, {Thalmann},
  {Beuzit}, {Dohlen}, {Feldt}, {Gratton}, {Kasper}, {Mouillet}, {Pragt},
  {Puget}, {Rigal}, {Roelfsema}, {Turatto}, {Udry}, {Waters}, \&
  {Wildi}}]{joos2011}
{Joos}, F., {Schmid}, H.~M., {Gisler}, D., {et~al.} 2011, in Astronomical
  Society of the Pacific Conference Series, Vol. 449, Astronomical Society of
  the Pacific Conference Series, ed. P.~{Bastien}, N.~{Manset}, D.~P.
  {Clemens}, \& N.~{St-Louis}, 381

\bibitem[{{Kasdin} {et~al.}(2003){Kasdin}, {Vanderbei}, {Spergel}, \&
  {Littman}}]{kasdin2003}
{Kasdin}, N.~J., {Vanderbei}, R.~J., {Spergel}, D.~N., \& {Littman}, M.~G.
  2003, \apj, 582, 1147

\bibitem[{{Kurucz}(1979)}]{kurucz1979}
{Kurucz}, R.~L. 1979, \apjs, 40, 1

\bibitem[{{L{\'e}na} {et~al.}(1998){L{\'e}na}, {Lebrun}, \&
  {Mignard}}]{lena1998}
{L{\'e}na}, P., {Lebrun}, F., \& {Mignard}, F. 1998, {Observational
  astrophysics.} ({Springer, Berlin (Germany)})

\bibitem[{{L{\'o}pez} {et~al.}(2007){L{\'o}pez}, {Bringas}, {Cuevas},
  {D{\'{\i}}az}, {Eikenberry}, {Espejo}, {Flores}, {Fuentes}, {Gallego},
  {Garz{\'o}n}, {Hammersley}, {Pell{\'o}}, {Prieto}, {S{\'a}nchez}, \&
  {Watson}}]{lopez2007}
{L{\'o}pez}, J.~A., {Bringas}, V., {Cuevas}, S., {et~al.} 2007, in Revista
  Mexicana de Astronomia y Astrofisica, Vol.~29, Revista Mexicana de Astronomia
  y Astrofisica Conference Series, 18--20

\bibitem[{{Macintosh} {et~al.}(2008){Macintosh}, {Graham}, {Palmer}, {Doyon},
  {Dunn}, {Gavel}, {Larkin}, {Oppenheimer}, {Saddlemyer}, {Sivaramakrishnan},
  {Wallace}, {Bauman}, {Erickson}, {Marois}, {Poyneer}, \&
  {Soummer}}]{macintosh2008}
{Macintosh}, B.~A., {Graham}, J.~R., {Palmer}, D.~W., {et~al.} 2008, in {SPIE
  Conference Series}, Vol. 7015

\bibitem[{{N'Diaye} {et~al.}(2008){N'Diaye}, {Cuevas}, \&
  {Dohlen}}]{ndiaye2008}
{N'Diaye}, M., {Cuevas}, S., \& {Dohlen}, K. 2008, in Society of Photo-Optical
  Instrumentation Engineers (SPIE) Conference Series, Vol. 7014, Society of
  Photo-Optical Instrumentation Engineers (SPIE) Conference Series

\bibitem[{{N'Diaye} {et~al.}(2007){N'Diaye}, {Dohlen}, \&
  {Cuevas}}]{ndiaye2007}
{N'Diaye}, M., {Dohlen}, K., \& {Cuevas}, S. 2007, in Proceedings of the
  conference In the Spirit of Bernard Lyot: The Direct Detection of Planets and
  Circumstellar Disks in the 21st Century. June 04 - 08, 2007. University of
  California, Berkeley, CA, USA. Edited by Paul Kalas., ed. P.~{Kalas}

\bibitem[{{N'Diaye} {et~al.}(2012){N'Diaye}, {Dohlen}, {Cuevas}, {Soummer},
  {S{\'a}nchez-P{\'e}rez}, \& {Zamkotsian}}]{ndiaye2012}
{N'Diaye}, M., {Dohlen}, K., {Cuevas}, S., {et~al.} 2012, \aap, 538, A55

\bibitem[{{Nisenson} \& {Papaliolios}(2001)}]{nisenson2001}
{Nisenson}, P. \& {Papaliolios}, C. 2001, \apjl, 548, L201

\bibitem[{{Patience} {et~al.}(2012){Patience}, {King}, {De Rosa}, {Vigan},
  {Witte}, {Rice}, {Helling}, \& {Hauschildt}}]{patience2012}
{Patience}, J., {King}, R.~R., {De Rosa}, R.~J., {et~al.} 2012, \aap, 540, A85

\bibitem[{{Petit} {et~al.}(2008){Petit}, {Fusco}, {Charton}, {Mouillet},
  {Rabou}, {Buey}, {Rousset}, {Sauvage}, {Baudoz}, {Gigan}, {Kasper},
  {Fedrigo}, {Hubin}, {Feautrier}, {Beuzit}, \& {Puget}}]{petit2008}
{Petit}, C., {Fusco}, T., {Charton}, J., {et~al.} 2008, in SPIE Conference
  Series, Vol. 7015

\bibitem[{{Pueyo} {et~al.}(2011){Pueyo}, {Crepp}, {Vasisht}, {Brenner},
  {Oppenheimer}, {Zimmerman}, {Hinkley}, {Parry}, {Beichman}, {Hillenbrand},
  {Roberts}, {Dekany}, {Shao}, {Burruss}, {Bouchez}, {Roberts}, \&
  {Soummer}}]{pueyo2011}
{Pueyo}, L., {Crepp}, J.~R., {Vasisht}, G., {et~al.} 2011, ArXiv e-prints

\bibitem[{{Soummer} {et~al.}(2007){Soummer}, {Pueyo}, {Sivaramakrishnan}, \&
  {Vanderbei}}]{soummer2007b}
{Soummer}, R., {Pueyo}, L., {Sivaramakrishnan}, A., \& {Vanderbei}, R.~J. 2007,
  Optics Express, 15, 15935

\bibitem[{{Sparks} \& {Ford}(2002)}]{sparks2002}
{Sparks}, W.~B. \& {Ford}, H.~C. 2002, \apj, 578, 543

\bibitem[{{Stahl} \& {Sandler}(1995)}]{stahl1995}
{Stahl}, S.~M. \& {Sandler}, D.~G. 1995, \apjl, 454, L153+

\bibitem[{{Thatte} {et~al.}(2007){Thatte}, {Abuter}, {Tecza}, {Nielsen},
  {Clarke}, \& {Close}}]{thatte2007}
{Thatte}, N., {Abuter}, R., {Tecza}, M., {et~al.} 2007, \mnras, 378, 1229

\bibitem[{{Vanderbei} {et~al.}(2004){Vanderbei}, {Kasdin}, \&
  {Spergel}}]{vanderbei2004}
{Vanderbei}, R.~J., {Kasdin}, N.~J., \& {Spergel}, D.~N. 2004, \apj, 615, 555

\bibitem[{{Vanderbei} {et~al.}(2003{\natexlab{a}}){Vanderbei}, {Spergel}, \&
  {Kasdin}}]{vanderbei2003b}
{Vanderbei}, R.~J., {Spergel}, D.~N., \& {Kasdin}, N.~J. 2003{\natexlab{a}},
  \apj, 599, 686

\bibitem[{{Vanderbei} {et~al.}(2003{\natexlab{b}}){Vanderbei}, {Spergel}, \&
  {Kasdin}}]{vanderbei2003}
{Vanderbei}, R.~J., {Spergel}, D.~N., \& {Kasdin}, N.~J. 2003{\natexlab{b}},
  \apj, 590, 593

\bibitem[{{V{\'e}rinaud} {et~al.}(2010){V{\'e}rinaud}, {Kasper}, {Beuzit},
  {Gratton}, {Mesa}, {Aller-Carpentier}, {Fedrigo}, {Abe}, {Baudoz},
  {Boccaletti}, {Bonavita}, {Dohlen}, {Hubin}, {Kerber}, {Korkiakoski},
  {Antichi}, {Martinez}, {Rabou}, {Roelfsema}, {Schmid}, {Thatte}, {Salter},
  {Tecza}, {Venema}, {Hanenburg}, {Jager}, {Yaitskova}, {Preis}, {Orecchia}, \&
  {Stadler}}]{verinaud2010}
{V{\'e}rinaud}, C., {Kasper}, M., {Beuzit}, J.-L., {et~al.} 2010, in SPIE
  Conference Series, Vol. 7736

\bibitem[{{Vigan} {et~al.}(2012{\natexlab{a}}){Vigan}, {Bonnefoy}, {Chauvin},
  {Moutou}, \& {Montagnier}}]{vigan2012}
{Vigan}, A., {Bonnefoy}, M., {Chauvin}, G., {Moutou}, C., \& {Montagnier}, G.
  2012{\natexlab{a}}, \aap, 540, A131

\bibitem[{{Vigan} {et~al.}(2008){Vigan}, {Langlois}, {Moutou}, \&
  {Dohlen}}]{vigan2008}
{Vigan}, A., {Langlois}, M., {Moutou}, C., \& {Dohlen}, K. 2008, \aap, 489,
  1345

\bibitem[{{Vigan} {et~al.}(2012{\natexlab{b}}){Vigan}, {Patience}, {Marois},
  {Bonavita}, {De Rosa}, {Macintosh}, {Song}, {Doyon}, {Zuckerman},
  {Lafreni{\`e}re}, \& {Barman}}]{vigan2012b}
{Vigan}, A., {Patience}, J., {Marois}, C., {et~al.} 2012{\natexlab{b}}, \aap,
  544, A9

\end{thebibliography}

\end{document}